\documentclass[twocolumn]{revtex4-1}
\usepackage[utf8]{inputenc}
\usepackage{braket}
\usepackage{amsmath}
\usepackage{amssymb}
\usepackage{graphicx}
\usepackage{subfigure}
\usepackage{float}
\usepackage{cancel}
\usepackage[colorlinks, citecolor=blue, linkcolor=blue, urlcolor=blue]{hyperref}

\DeclareMathOperator*{\argmin}{arg\,min}
\DeclareMathOperator{\sign}{sign}
\newcommand{\dmct}{\delta\tau}

\begin{document}
\title{Stochastic nodal surfaces in quantum Monte Carlo calculations}

\author{Michael Hutcheon}
\email{mjh261@cam.ac.uk}
\affiliation
{
    Theory of Condensed Matter Group,
    Cavendish Laboratory,
    J.J.~Thomson Avenue,
    Cambridge CB3 0HE,
    United Kingdom
}

\date{\today}

\begin{abstract}
    Treating the fermionic ground state problem as a constrained stochastic optimization problem, a formalism for fermionic quantum Monte Carlo is developed that makes no reference to a trial wavefunction. Exchange symmetry is enforced by nonlocal terms appearing in the Green's function corresponding to a new kind of walker propagation. Complemented by a treatment of diffusion that encourages the formation of a stochastic nodal surface, we find that an approximate long-range extension of walker cancellations can be employed without introducing significant bias, reducing the number of walkers required for a stable calculation. A proof-of-concept implementation is shown to give a stable fermionic ground state for simple harmonic and atomic systems.
\end{abstract}

\maketitle

Quantum Monte Carlo (QMC) methods have provided some of the most important results in computational physics \cite{ceperley_alder} and remain amongst the most accurate methods available for calculating ground state properties of quantum systems \cite{qmc_accuracy}. However, for certain systems, QMC suffers from the infamous \textit{fermion sign problem}, the general solution to which has been shown to be NP-hard \cite{qmc_np_hard}. We focus on the specific case of \textit{diffusion Monte Carlo} (DMC) methods \cite{qmc_needs, qmc_austin, qmc_toulouse}, which converge on the many-body ground state by iteratively projecting out exited state components from the wavefunction. Here, the sign problem arises due to exchange symmetry dividing the wavefunctions into regions of different sign, known as \textit{nodal pockets}, separated by a \textit{nodal surface}. This increases the fermionic ground state energy relative to that of the bosonic ground state and, as a result, the former is projected out, typically exponentially decaying away as the iterative procedure progresses \cite{fermion_monte_carlo_revisited}.

Despite this exponential decay of the fermionic component, methods such as \textit{release-node} DMC can extract information about the fermionic ground state from the transient behaviour of the wavefunction \cite{release_node_ceperly}. However, this transient behaviour leads to a statistical error that grows with system size, requiring a formidable computational effort to mitigate \cite{prospects_for_release_node}. The most popular approach to obtain a stable fermionic ground state in DMC is known as the \textit{fixed-node approximation}, developed in the early 80s \cite{early_fixed_node, early_fixed_node_2}, whereby the nodal surface is fixed to that of some trial wavefunction, which must be known \textit{a priori}. We focus on electronic systems, where it is conjectured that the presence of many-body correlation leads to the minimal case of only two nodal pockets \cite{only_two_nodal_cells}, which may make the electronic problem more tractable than the general NP-hard case.

In this work, we develop a formalism of fermionic DMC that makes no reference to a trial wavefunction. In section \ref{sec:formalism} we show that including exchange symmetry as a constraint in the energy minimization problem leads to a modified DMC scheme, resulting in a new propagation channel in the Green's function that couples populations of signed walkers. In section \ref{sec:implementation} we go on to show how this propagation results in the formation of a \textit{stochastic} nodal surface, constructed from the entire population of walkers, that is free to vary and minimize the energy. We propose a diffusion scheme to maximise it's stability. 

Compared to previous methods employing signed walkers (such as \textit{the fermion Monte Carlo} method of Kalos and Pederiva \cite{kalos_fermion_monte_carlo}, the early work of Anderson and Traynor \cite{quantum_chemistry_by_random_walk_exact} or the second-quantized approach of Umrigar \cite{umrigar_observations}) our work presents an alternative way to manage the propagation, and cancellations between, signed walkers, informed by the underlying optimization problem. In section \ref{sec:discussion} we discuss the relationship to these previous schemes, which themselves can be thought of as having fluctuating nodal surfaces due to walker-walker cancellation processes. Finally, we provide an open-source implementation of the method \cite{github} and demonstrate that it obtains a stable fermionic ground state for the harmonic and atomic systems considered.

\section{Formalism}
\label{sec:formalism}

We start by formulating the fermionic problem for $N$ particles in $d$ dimensions as the following \textit{constrained} optimization problem:

\begin{equation*}
    \text{Find} \;  \argmin_{\ket{\psi}}\bra{\psi} H \ket{\psi} \; \text{such that}
\end{equation*}
\vspace{-0.5cm}
\begin{align}
    \langle \psi \ket{\psi} &= 1 &\text{(Normalization)} \label{eq:normalization_constraint} \\
    \langle x \ket{\psi} &= - \langle P_i x \ket{\psi} &\text{(Antisymmetry)}
    \label{eq:antisymmetry_constraint}
\end{align}
\vspace{-0.5cm}
\begin{equation*}
    \forall \; P_i \in \mathcal{E},\; x \in \mathbb{R}^{dN}
\end{equation*}
where $\mathcal{E}$ is the set of pairwise identical-fermion exchanges. If the system contains $M$ identical fermions, there are $M(M-1)/2$ such exchanges. These exchanges can be combined to generate the set $\mathcal{P}$ of the $M!$ permutations of identical fermions. Introducing the Lagrange multipliers $E_T$ and $\mu_i(x)$ the optimization problem is equivalent to extremizing the Lagrangian
\begin{multline}
    \mathcal{L} = \bra{\psi} H \ket{\psi} + E_T\left[1 - \langle\psi\ket{\psi} \right] \\ + \sum_i \int \psi^*(x)
    \mu_i(x)(P_i + 1)\psi(x) dx
\end{multline}
with respect to $\psi$, $\psi^*$, $E_T$ and the $\mu_i(x)$'s. We note that $\mathcal{L}$ can be written as
\begin{equation}
    \mathcal{L} = E_T + \bra{\psi} \overbrace{H - E_T + \sum_i \mu_i(x)(P_i + 1)}^{H_X} \ket{\psi}
\end{equation}
allowing us to define an effective Hamiltonian $H_X$. The term involving the Lagrange multipliers $\mu_i$ can be interpreted as a cost function that penalises the appearance of a symmetric component in the wavefunction so long as we require $\mu_i(x) > 0$. Extremization of $\mathcal{L}$ with respect to $\psi$ and $\psi^*$ \cite{supplement} leads to
\begin{equation}
    H_X\psi = 0 = H\psi - E_T\psi + \left[\sum_i \mu_i(x)(P_i + 1)\right]\psi
\end{equation}
From Eq.\ \ref{eq:antisymmetry_constraint} we see that the term in square brackets vanishes at the extremum of $\mathcal{L}$, leading to the Schr\"odinger equation $H\psi = E_T\psi$. This allows us to identify $E_T$ as the fermionic ground state energy. 

To perform the extremization we propagate the imaginary time ($\tau = it$) Schr\"odinger equation for $H_X$,
\begin{equation}
\label{eq:imaginary_time_schrodinger}
    \frac{\partial \ket{\psi(\tau)}}{\partial \tau} = - H_X \ket{\psi(\tau)}
\end{equation}
which can be written in integral form as
\begin{equation}
\label{eq:imaginary_time_propagation}
    \underbrace{\langle x\ket{\psi(\tau+\dmct)}}_{\underset{\psi(x,\tau+\dmct)}{\text{Propagated wavefunction}}} \hspace{-0.5cm} = \hspace{-0.1cm}
    \int \hspace{-0.1cm} \underbrace{\bra{x} \exp(-\dmct H_X) \ket{x'}}_{\underset{G(x,x',\dmct)}{\text{Green's function}}} \hspace{-0.3cm}
    \underbrace{\langle x' \ket{\psi(\tau)}}_{\underset{\psi(x',\tau)}{\text{Old wavefunction}}} \hspace{-0.3cm} dx'.
\end{equation}
Following traditional DMC, we sample our wavefunction with a discrete set of walkers, each representing a particular point in configuration space $x_i$ and carrying a corresponding weight $w_i$:
\begin{equation}
    \psi_{\text{DMC}}(x, \tau) = \sum_i w_i(\tau) \delta(x - x_i(\tau)).
\end{equation}
Eq.\ \ref{eq:imaginary_time_propagation} can then be interpreted as an evolution equation for the walkers, where the Green's function $G(x,x',\dmct)$ enters as a generalised transition probability from $x' \rightarrow x$. Substituting $\psi_{\text{DMC}}$ into this evolution equation, we obtain the  propagated wavefunction
\begin{equation}
\label{eq:delta_to_greens_function}
    \psi_{\text{DMC}}(x, \tau+\dmct) = \sum_i w_i(\tau) G(x, x_i(\tau), \dmct).
\end{equation}
Writing $H = T + V$, where $T$ is the kinetic energy operator and $V$ is the (local) many-body potential, allows us to define the well-known \cite{qmc_needs} potential and diffusive parts of the Green's function
\begin{align*}
    G_V(x, x', \dmct) &\equiv \exp\left(-\dmct[V(x)+V(x')]/2\right),\\
    G_D(x,x',\dmct) &\equiv \bra{x}\exp(-\dmct T)\ket{x'} \propto \exp \left(-\frac{|x-x'|^2}{2\dmct}\right).
\end{align*}
For sufficiently small timesteps $\dmct \ll 1$, our full Green's function can then be written (see appendix \ref{app:greens_function}) as
\begin{equation}
\begin{aligned}
\label{eq:greens_function_full}
    G(x,x',\dmct) &= \overbrace{\left[\mathcal{N}(x') - \sum_i \mathcal{X}_i(x') P_i\right]}^{\substack{G_X \Longleftrightarrow\; \text{Exchange moves}}} \\ \times \underbrace{G_V(x,x',\dmct)}_{\text{Potential weighting}} \hspace{-0.3cm} &\times \underbrace{G_D(x,x',\dmct)}_{\text{Diffusion}} \times \hspace{-0.3cm} \underbrace{\exp(\dmct E_T)}_{\text{Population control}},
\end{aligned}
\end{equation}
with
\begin{equation}
\begin{aligned}
    \mathcal{X}_i(x') &= \dmct\mu_i(P_ix'), \\
    \mathcal{N}(x') &= 1 - \sum_i \dmct \mu_i(x').
\end{aligned}
\end{equation}

\begin{figure}
    \centering
    \includegraphics[width=\columnwidth]{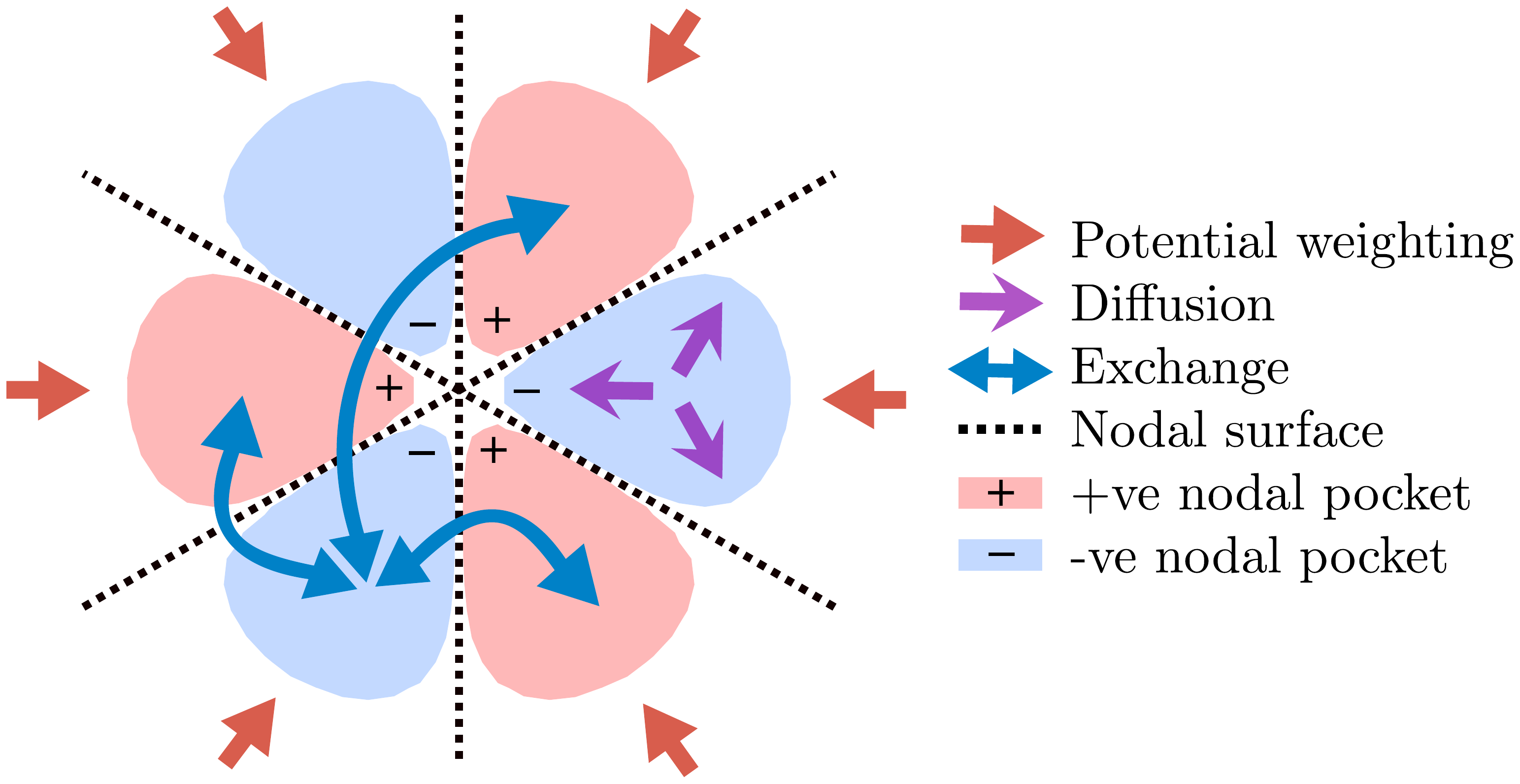}
    \caption{Schematic of wavefunction formation arising from competing walker propagation channels (shown for three fermions in a harmonic well as in Fig.\ \ref{fig:3nif}).}
    \label{fig:wavefunction_formation}
\end{figure}

We note that if we were to neglect the fermionic constraint, we would recover the Green's function of traditional DMC \cite{qmc_needs}. The part arising from this constraint is labelled $G_X$ and can be applied to a walker at $x'$ with weight $w$ by carrying out the fermionic exchange $\{x' \rightarrow P_ix', w\rightarrow -w\}$ with probability $\mathcal{X}_i(x')$. These non-local \textit{exchange moves} enforce the antisymmetry of the wavefunction by allowing walkers sampling one nodal pocket to stochastically switch to sampling any symmetry-related nodal pocket (see Fig.\ \ref{fig:wavefunction_formation}). The tiling theorem \cite{ceperley_fermion_nodes} then implies that any walker can access and contribute weight to all nodal pockets. As a result, rather than each walker simply contributing to the wavefunction at a particular point in configuration space, it can now contribute to all symmetry-related points. For simplicity, in our implementation we choose the probabilities $\mathcal{X}_i(x')$ and $\mathcal{N}(x')$ so that each of the exchange moves (including no exchange) are equiprobable. 

\section{Implementation}
\label{sec:implementation}

\subsection{Stochastic nodal surface}
\label{sec:stochastic_nodal_surface}

\begin{figure}
    \centering
    \includegraphics[width=0.8\columnwidth]{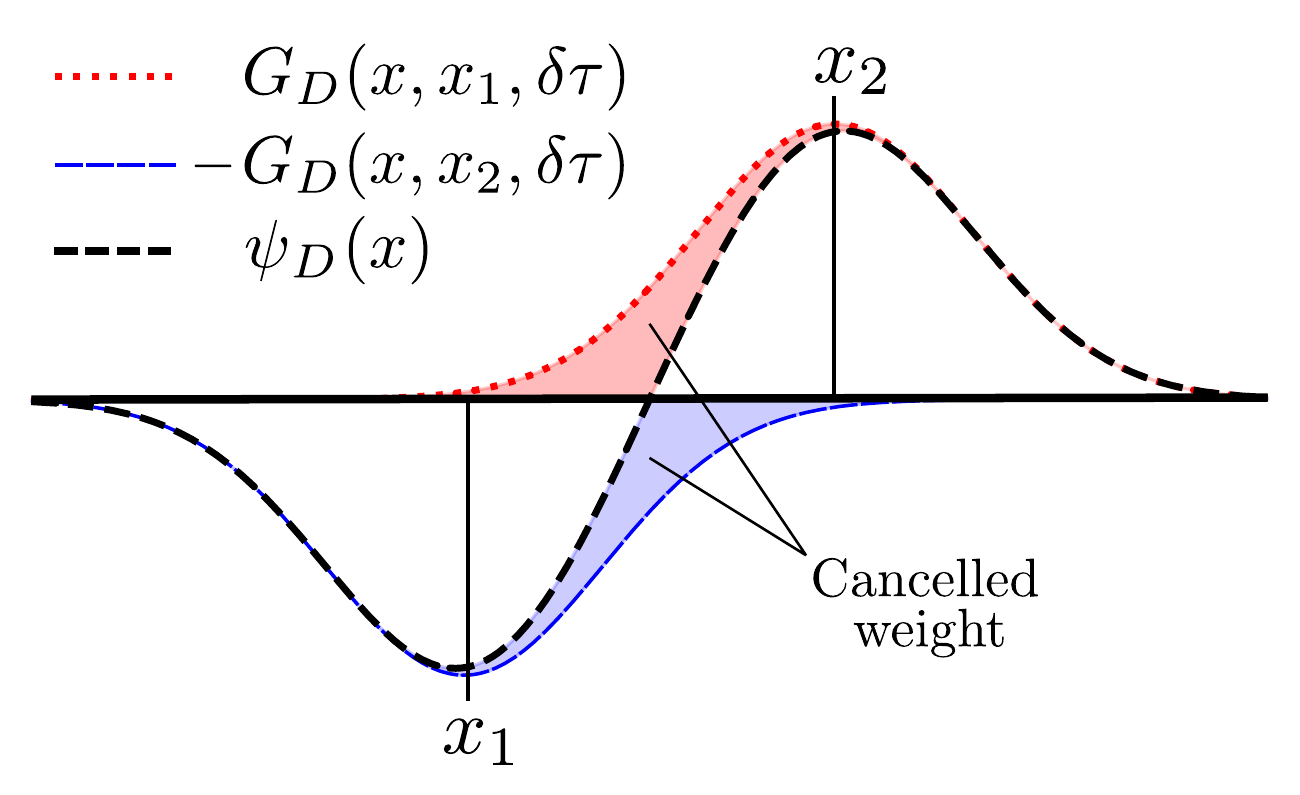}
    \caption{The diffusive propagation of two nearby walkers of opposite sign located at $x_1$ and $x_2$ $\implies$ $\psi_D(x) = G_D(x, x_2, \dmct) - G_D(x, x_1, \dmct) \equiv G_2 - G_1$ (black dashed line). The red (blue) shaded region show the portion of $G_2$ ($G_1$) that can be cancelled in the propagation.}
    \label{fig:dmc_prop_wfn}
    \vspace{0.25cm}
    \centering
    \includegraphics[width=0.8\columnwidth]{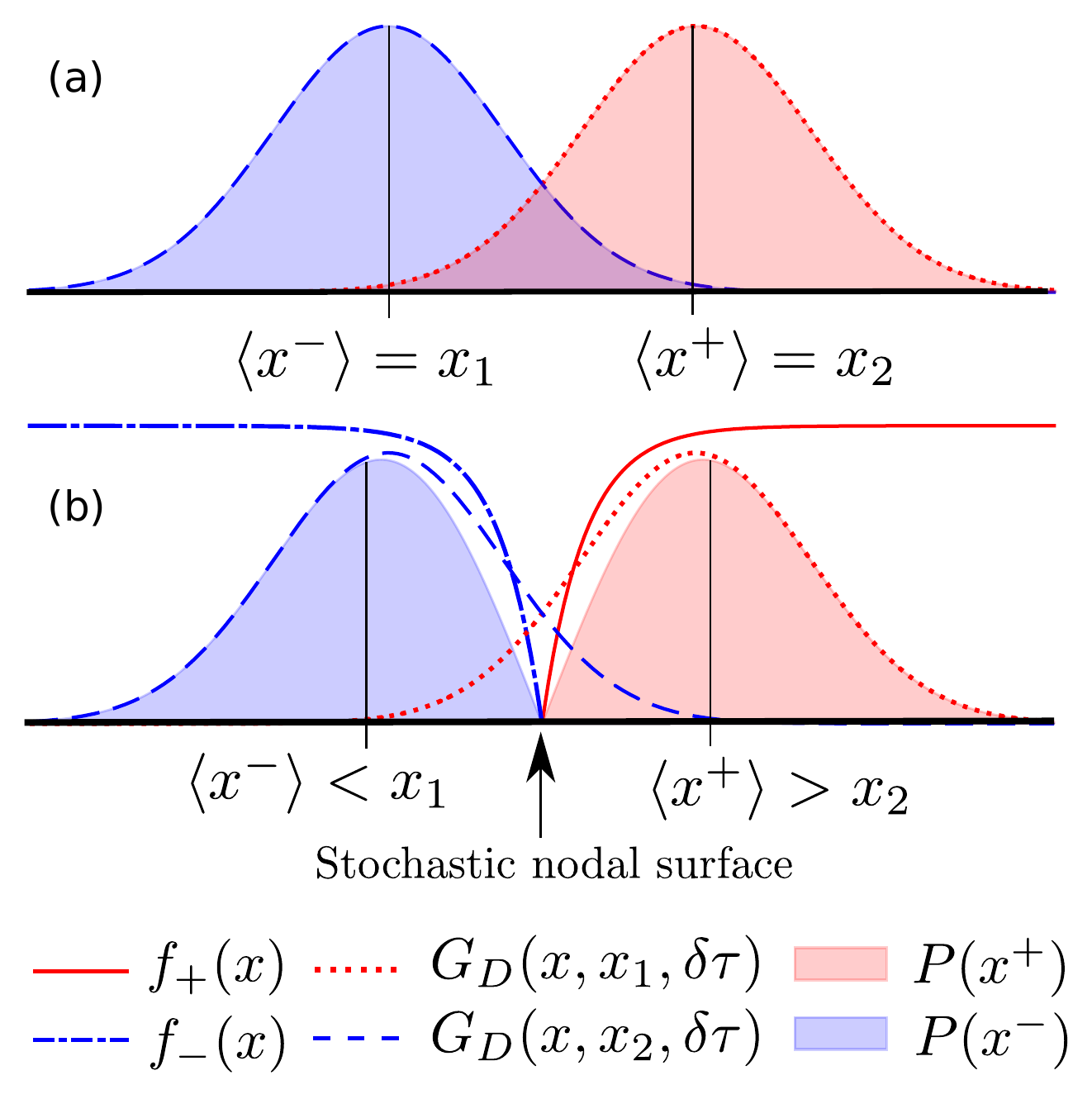}
    \caption{Propagation schemes satisfying Eq.\ \ref{eq:propagation_condition}, applied to the walkers in Fig.\ \ref{fig:dmc_prop_wfn}. (a) traditional DMC propagation (Eq.\ \ref{eq:traditional_dmc_propagation}). (b) our propagation scheme (Eq.\ \ref{eq:prop_scheme}). Note that in (a) there is overlap of the +ve and -ve walker distributions. The same is not true for (b).}
    \label{fig:dmc_prop_schemes}
\end{figure}

\begin{figure}
    \centering
    \includegraphics[width=0.9\columnwidth]{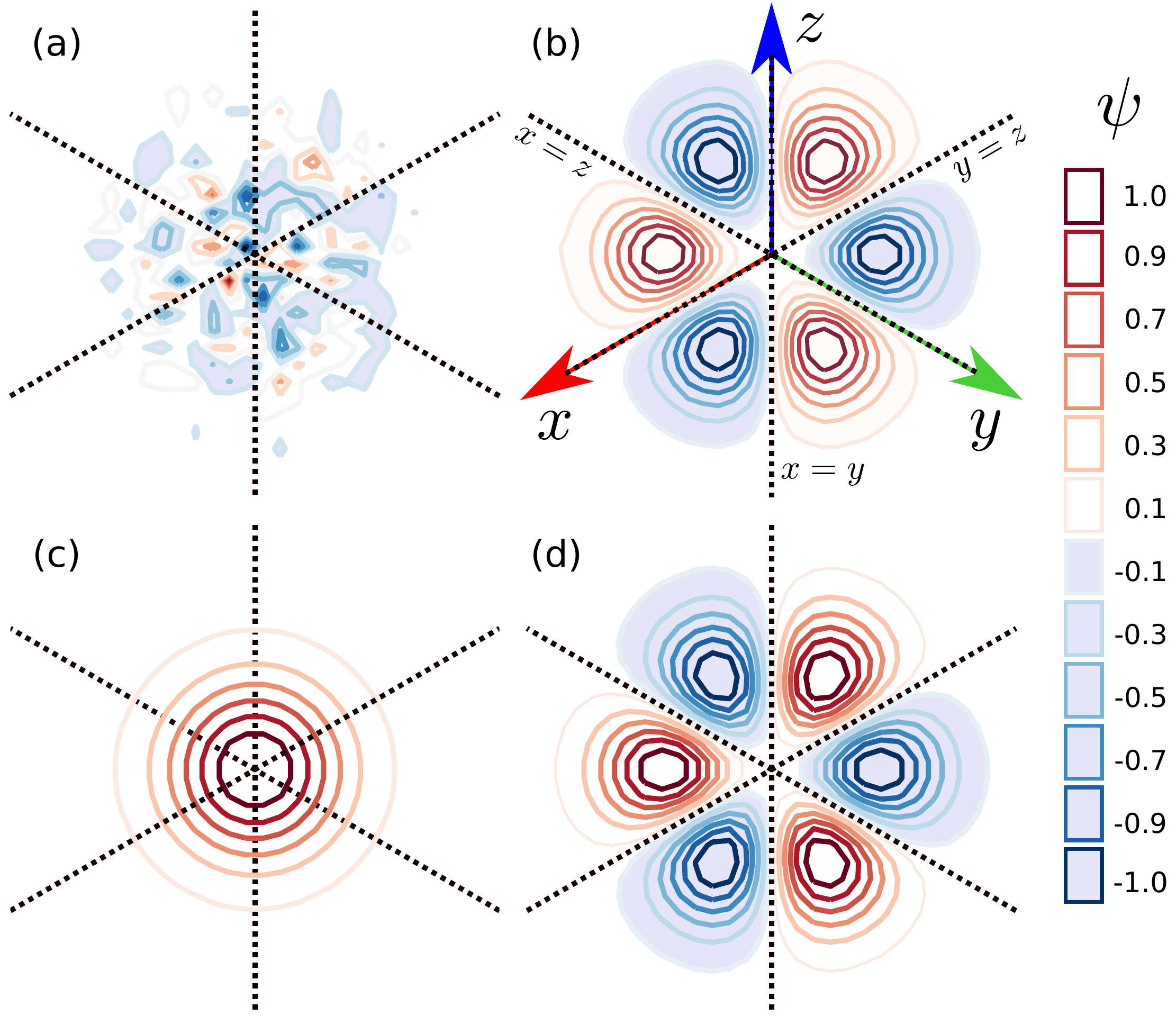}
    \caption{The wavefunction of three non-interacting fermions with coordinates $x,\;y$ and $z$ in a one-dimensional harmonic well, integrated and viewed along the $(1,1,1)$ projection. The analytic nodal surface is shown as a dotted black line. From this projection, the nodal pockets can be clearly seen. (a) Bosonic collapse from DMC with exchange moves but without a stochastic nodal surface. (b) From DMC with exchange moves and a stochastic nodal surface. (c) Analytic bosonic ground state. (d) Analytic fermionic ground state.}
    \label{fig:3nif}
\end{figure}

To maximise the effectiveness of the exchange moves, we also consider how best to apply the other parts of the Green's function. The diffusive part of the Green's function applied to a set of walkers leads to the diffused wavefunction
\begin{equation}
    \psi_D(x) = \sum_i w_i G_D(x,x_i,\dmct)
\end{equation}
as shown in Fig.\ \ref{fig:dmc_prop_wfn} for two opposite-sign walkers. If we represent this new wavefunction as a combination of walkers with weights $\pm1$ with configurations sampled from the distributions $P_\pm(x)$ respectively, we must have
\begin{equation}
\label{eq:propagation_condition}
    P_+(x) - P_-(x) = \psi_D(x).
\end{equation}
In traditional DMC each walker diffuses independently by an amount sampled from $G_D$, resulting in
\begin{equation}
\begin{aligned}
\label{eq:traditional_dmc_propagation}
    P_+(x) &= \psi_+(x) \equiv \sum_{w_i > 0} w_i(\tau) G_D(x,x_i,\dmct) \\
    P_-(x) &= \psi_-(x) \equiv \sum_{w_i < 0} |w_i(\tau)| G_D(x,x_i,\dmct).
\end{aligned}
\end{equation}
A drawback of this scheme when applied to signed walkers is that it allows +ve walkers to move into a region where $\psi_D$ is -ve, and vice versa, as can be seen from the overlap of $P_+(x)$ and $P_-(x)$ in Fig.\ \ref{fig:dmc_prop_schemes}(a). This prohibits the emergence of well-separated regions of +ve and -ve walkers, corresponding to nodal pockets. Without stable nodal pockets, the walkers end up sampling the bosonic ground state with a randomly fluctuating sign. This is known as \textit{bosonic collapse} and arises in a similar fashion to the exponentially decaying signal-to-noise ratio in so-called \textit{release-node} DMC \cite{fermion_monte_carlo_revisited}. An example is shown in Fig.\ \ref{fig:3nif}(a) for a system of three non-interacting fermions in a harmonic well.

To avoid bosonic collapse, one particular sign of walker should dominate at each point in configuration space. Typically this sign is chosen according to the fixed-node approximation as being equal to that of the trial wavefunction. We instead derive a propagation scheme that encourages the formation of a \textit{stochastic} nodal surface which, in contrast to fixed-node DMC, is free to vary and minimize the energy. In order to encourage the formation of such a nodal surface, we seek the form of $P_\pm(x)$ that maximizes the expected separation of +ve and -ve walkers, given by
\begin{equation}
    \langle|x_+ - x_-|\rangle = \int P_+(x_+)P_-(x_-)|x_+ - x_-| dx_+ dx_-
\end{equation}
This leads, independently of the form of $\psi_D(x)$ (see appendix \ref{app:prop_scheme}), to
\begin{equation}
    P_\pm(x) = 
    \begin{cases} 
          |\psi_D(x)| & \text{if}\; \sign(\psi_D(x)) = \pm 1, \\
          0          & \text{otherwise}.
    \end{cases} \label{eq:prop_scheme}
\end{equation}
These distributions have no overlap, as can be seen in Fig.\ \ref{fig:dmc_prop_schemes}(b). However, because $P_\pm(x)$ are no longer simple sums of Gaussian terms (as $\psi_\pm$ were), they are difficult to sample moves from directly. This can be remedied by noting that
\begin{equation}
\label{eq:propagation_factorisation}
    P_\pm(x) = \psi_\pm(x)f_\pm(x)
\end{equation}
where \cite{supplement}
\begin{equation}
\label{eq:cancellation_function}
    f_\pm(x) = \max\left(1- \psi_\mp(x)/\psi_\pm(x),\; 0\right) \in [0,1]
\end{equation}
can be interpreted as reweighting functions, as shown in Fig.\ \ref{fig:dmc_prop_schemes}(b). We can then interpret Eq.\ \ref{eq:propagation_factorisation} as a diffusion according to $\psi_\pm(x)$ (corresponding to moves sampled from $G_D(x,x',\dmct)$) followed by a corrective reweighting $w\rightarrow f_\pm(x)w$, due to cancellation of +ve and -ve walkers.

Applying this scheme to the same system of three non-interacting fermions in a harmonic well results in the wavefunction shown in Fig.\ \ref{fig:3nif}(b). Comparing to Fig.\ \ref{fig:3nif}(d) we see that the analytic nodal surface is reproduced.

\subsection{Initialization}
\label{sec:initialization}
We initialize the walkers in such a way as to speed up their equilibration into an antisymmetric state. This is achieved by defining a unique ordering of the walker configurations, whereby the particles are ordered by their increasing coordinates. For example, in 2 spatial dimensions, the particles are ordered first by increasing $x$ coordinate, then by increasing $y$ coordinate. Starting with walker configurations distributed according to a normal distribution $x_i \sim \mathcal{N}(\mu=0, \sigma=1 \text{a.u})$, we apply exchange moves to the walkers $x = (\textbf{r}_1, \textbf{r}_2, ... ,\textbf{r}_N)$ until their constituent particles are increasing according to this order (i.e $\textbf{r}_1 \leq \textbf{r}_2 \leq \textbf{r}_3 ...$), and set their weights to +1. This is the same as the ordering used in Ref.\ \cite{grid_based_monte_carlo_2020}, except here we only use this procedure for initialization of the walkers. As the simulation proceeds, walkers will be propagated into antisymmetric images of this initial positive-definite group, quickly setting up a large antisymmetric component.

\subsection{Energy estimation}
\label{sec:energy_estimators}
The Lagrange multiplier associated with normalization, $E_T$, corresponds to an energy offset which appears in our effective Hamiltonian $H_X$. As the algorithm progresses, the value of $E_T$ is updated to keep the total weight of walkers, $W(\tau) = \sum_i |w_i(\tau)|$, roughly constant. The expected total weight after propagation from $\tau$ to $\tau+\dmct$ is given by
\begin{equation}
\begin{aligned}
    \langle W(\tau+\dmct) \rangle &= \sum_i | \langle w_i(\tau+\dmct) \rangle |\\
    &= \sum_i |w_i(\tau)G(x_i(\tau+\dmct), x_i(\tau), \dmct)|.
\end{aligned}
\end{equation}
Separating this into contributions from different parts of the Green's function (see Eq.\ \ref{eq:greens_function_full}) we have
\begin{equation}
     \langle W(\tau+\dmct) \rangle = \sum_i |w_i(\tau)G_X^{(i)}G_V^{(i)}G_D^{(i)}\exp(\delta\tau E_T)|
\end{equation}
where we have used the shorthand notation $G^{(i)} \equiv G(x_i(\tau+\dmct), x_i(\tau), \dmct)$. We keep the total weight roughly constant by requiring
\begin{equation}
\label{eq:growth_estimator}
\begin{aligned}
    \langle W(\tau+\dmct) \rangle &\overset{!}{=} W(\tau) = \sum_i |w_i(\tau)| \implies \\
    E_T(\tau) &= \frac{1}{\dmct} \ln\left(\frac{\sum_i |w_i(\tau)|}{\sum_i |w_i(\tau)G_X^{(i)}G_V^{(i)}G_D^{(i)}| }\right).
\end{aligned}
\end{equation}
This is known as the \textit{growth estimator} of the energy and, in order to keep the population stable, will converge to the lowest eigenvalue of $H_X$, which we can therefore estimate by averaging the value of $E_T(\tau)$ (after equilibration) over many iterations. However, because each iteration is correlated with the previous iteration, one must be careful in estimating the uncertainty of such averages. In this work we use the widely-employed reblocking method \cite{reblocking} to estimate statistical uncertainties.

\begin{figure}
    \centering
    \includegraphics[width=\columnwidth]{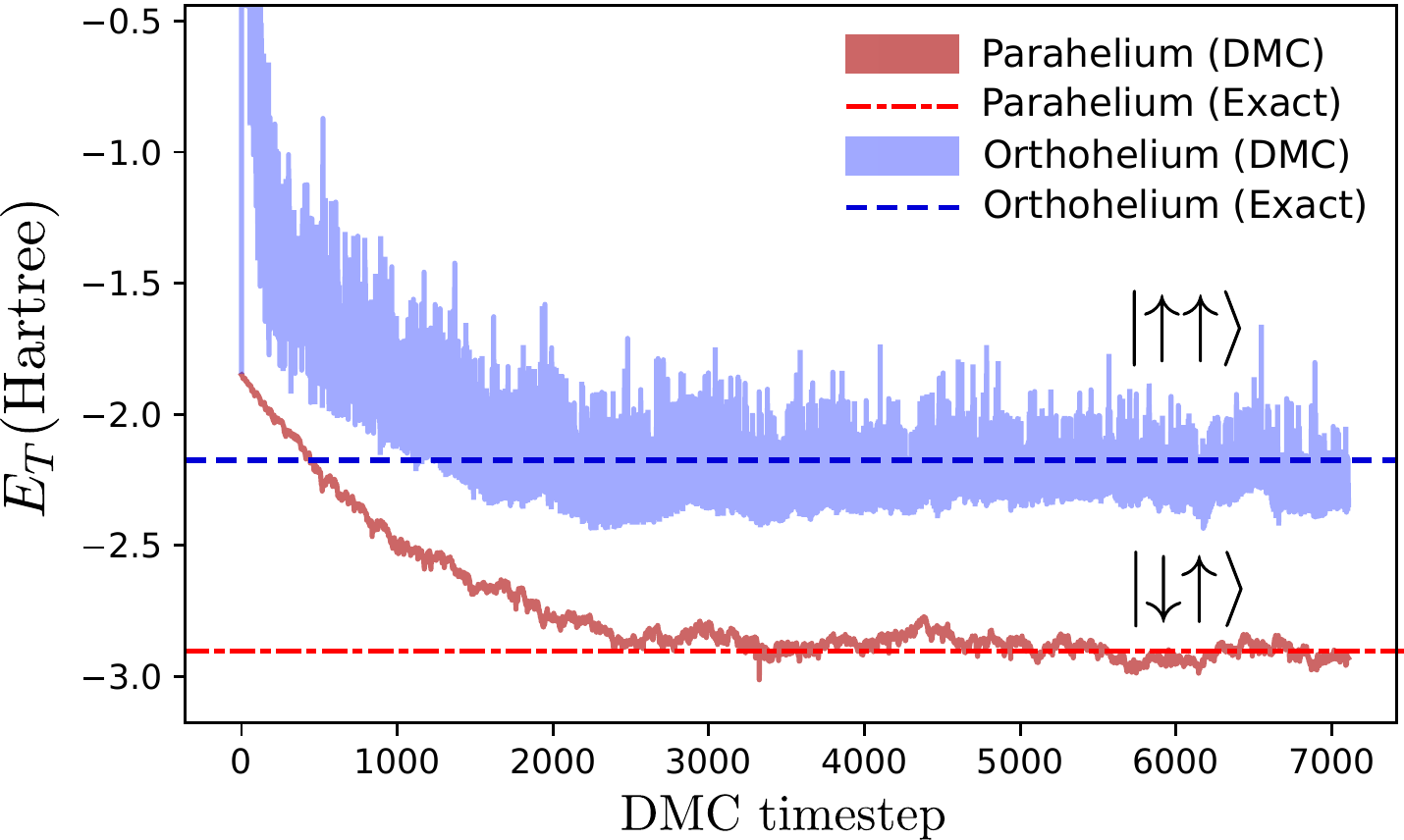}
    \caption{DMC calculations of the Helium atom ground state, with electrons having opposite spin (parahelium), and the exited (triplet) state with electrons having parallel spin (orthohelium), calculated using a stochastic nodal surface. A timestep of $\delta\tau=10^{-3}$ atomic units was used with $\delta\tau_{\text{eff}}$ = 0.5 atomic units (see section \ref{sec:tau_eff}). It is clear to see that the exited (fermionic) state shows larger fluctuations than the (bosonic) ground state. This is due to cancellations between oppositely signed walkers contributing to fluctuations in the growth estimator of the energy. The reference energies are from VMC optimization of many-parameter trial wavefunctions \cite{helium_energies}, accurate to within a few $\mu$Ha.}
    \label{fig:helium_exited_state}
\end{figure}

\begin{figure}
    \centering
    \includegraphics[width=\columnwidth]{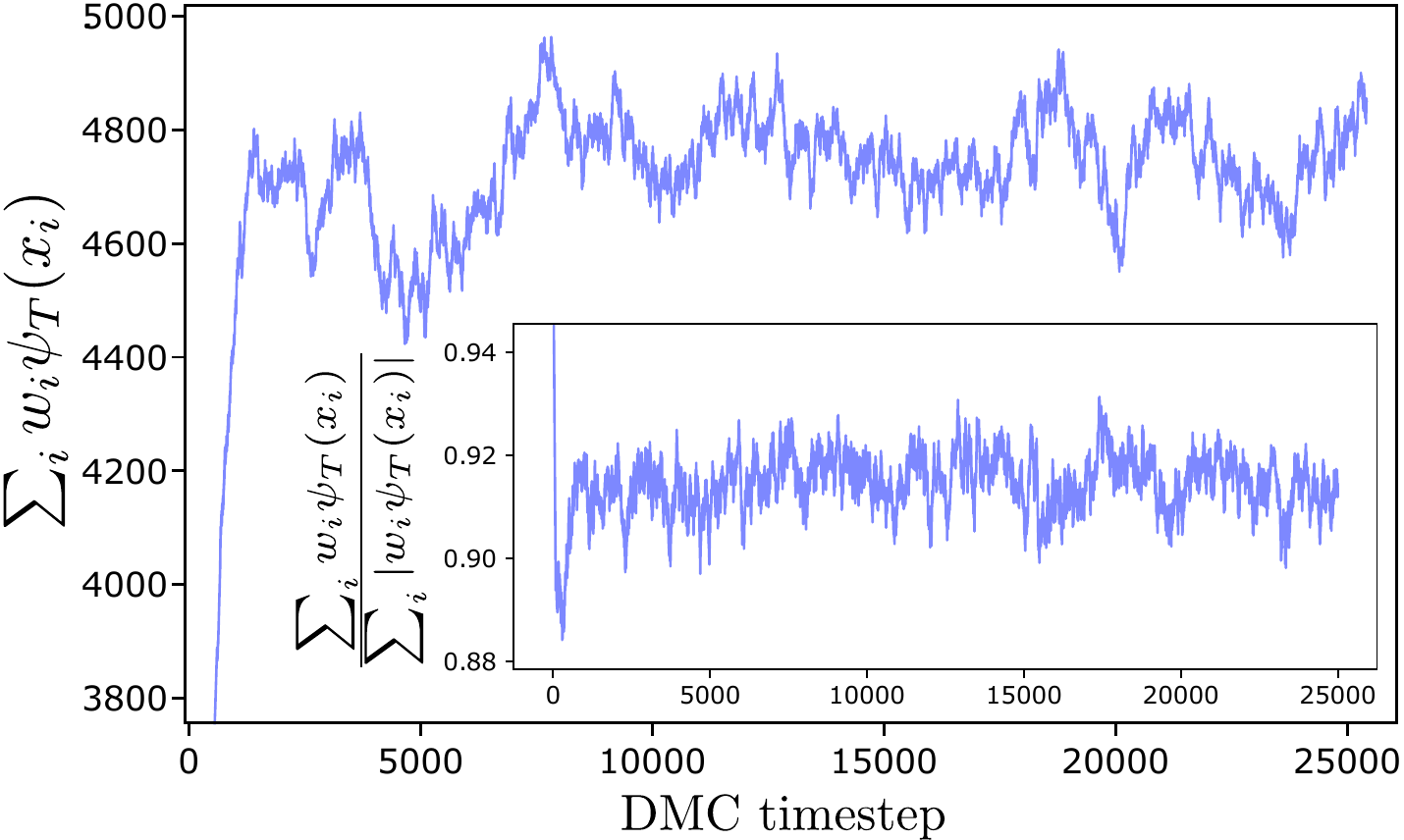}
    \caption{The denominator of Eq.\ \ref{eq:projection_estimator} vs. DMC timestep for the simulation used to produce Fig.\ \ref{fig:3nif}(b) ($10^4$ walkers, $\delta\tau = \delta\tau_{\text{eff}} = 10^{-3}$ atomic units). Note the y-axis scale. For the purposes of this plot, the trial wavefunction was set to the analytic fermionic ground state (shown in Fig.\ \ref{fig:3nif}(d)). We can see that the denominator remains large and roughly constant. Inset: the denominator as a fraction of it's maximum possible value (obtained if the walkers are all of the same sign as the analytic wavefunction).}
    \label{fig:3nif_overlap}
\end{figure}

If one has access to a suitable trial wavefuncion $\psi_T(x)$, that has non-zero overlap with the exact fermionic ground state, the fermionic energy can be estimated directly using the so-called projection estimator:
\begin{equation}
\label{eq:projection_estimator}
    E_{\text{proj}}(\tau) = \frac{\sum_i w_i(\tau) H \psi_T(x_i)}{\sum_i w_i(\tau) \psi_T(x_i)}
\end{equation}
Note that it is $H$, not $H_X$, that appears in Eq.\ \ref{eq:projection_estimator}. It is well-known \cite{grid_based_monte_carlo_2020} that the statistical uncertainties in $E_{\text{proj}}(\tau)$ are typically smaller than that of $E_T(\tau)$, due to the reduced dependence on the fluctuating population. Cancellations between signed walkers contribute to these fluctuations, resulting in larger statistical errors for fermionic systems when using the growth estimator. This can be clearly seen by comparing DMC calculations of the ground state and first exited state of the Helium atom in Fig.\ \ref{fig:helium_exited_state}. However, in order to use Eq.\ \ref{eq:projection_estimator}, it's denominator must remain finite for sufficiently many DMC timesteps to build up accurate statistical averages. In transient methods, such as \textit{release-node} DMC, the exponential decay of the fermionic component leads to an exponential decay of the denominator of Eq.\ \ref{eq:projection_estimator} and a correspondingly small set of usable iterations from which to build up such averages. In contrast, we find that the fermionic state obtained from propagating the Green's function of $H_X$ leads to stable (at least on the timescales we have probed in obtaining the results for this work) non-zero denominator of Eq.\ \ref{eq:projection_estimator}, as can be seen in Fig.\ \ref{fig:3nif_overlap}, allowing straightforward use of projection-based estimators. However, both for simplicity and as a proof-of-concept, we restrict ourselves to considering implementations that require no trial wavefunction, and so are limited to using the growth estimator. In doing this we are computing the `bosonic' (lowest) energy of the Greens function, rather than estimating just the fermionic component. As a result, any symmetric component that remains (despite the exchange-moves and cancellations) will influence the estimator towards the symmetric ground state energy (see for example Fig.\ \ref{fig:lithium}), rather than being removed by a projection-type estimator. The extension of the method to include a trial wavefunction to allow both the use of the projection estimator and importance sampling (see Ref.\ \cite{qmc_needs}) is a high priority for future work.

As is typical in DMC, after modifying the weights according to each part of the Green's function, we treat them with a \textit{birth-death} algorithm. This algorithm is designed to stop a single walker (usually in a low-potential region) simply accumulating all of the weight and exponentially dominating over the rest. In our implementation a walker with weight $w_i$ is replaced with $\lfloor |w_i| + u \rfloor$ walkers, each with weight $\sign(w_i)$. Here $u$ is a uniform random number $\in [0,1]$ and $\lfloor \cdot \rfloor$ is the floor function. This procedure leaves $\langle W \rangle$ unchanged, whilst preventing individual weights from becoming too small or large.

In atomic systems, timestep error can lead to a walker diffusing too close to a configuration where an electron overlaps with a nucleus and obtaining a correspondingly divergent (+ve) weight. This is known as a \textit{population explosion}. We mitigate this outcome by defining a maximum walker weight $w_{\text{max}}$ and reverting any DMC iteration where $\max(|w_i|) > w_{\text{max}}$ (in this work $w_{\text{max}} \geq 4$ resulting in only 1 in every $\sim$5000 iterations being reverted). We also use a softened version of the coulomb interaction of the form
\begin{equation}
    V_{c,\text{soft}}(r,r_s) = \frac{1}{r+r_s}
\end{equation}
For the calculations performed in this work $r_s \leq 10^{-5}$ which introduces a bias that is much smaller than the timestep error. We note that schemes to reduce the error due to coulomb singularities exist  \cite{coulomb_pseudopotential, qmc_attractive_coulomb}, but are not employed here.

\subsection{Effective nodal surface timestep}
\label{sec:tau_eff}

\begin{figure}
    \centering
    \includegraphics[width=\columnwidth]{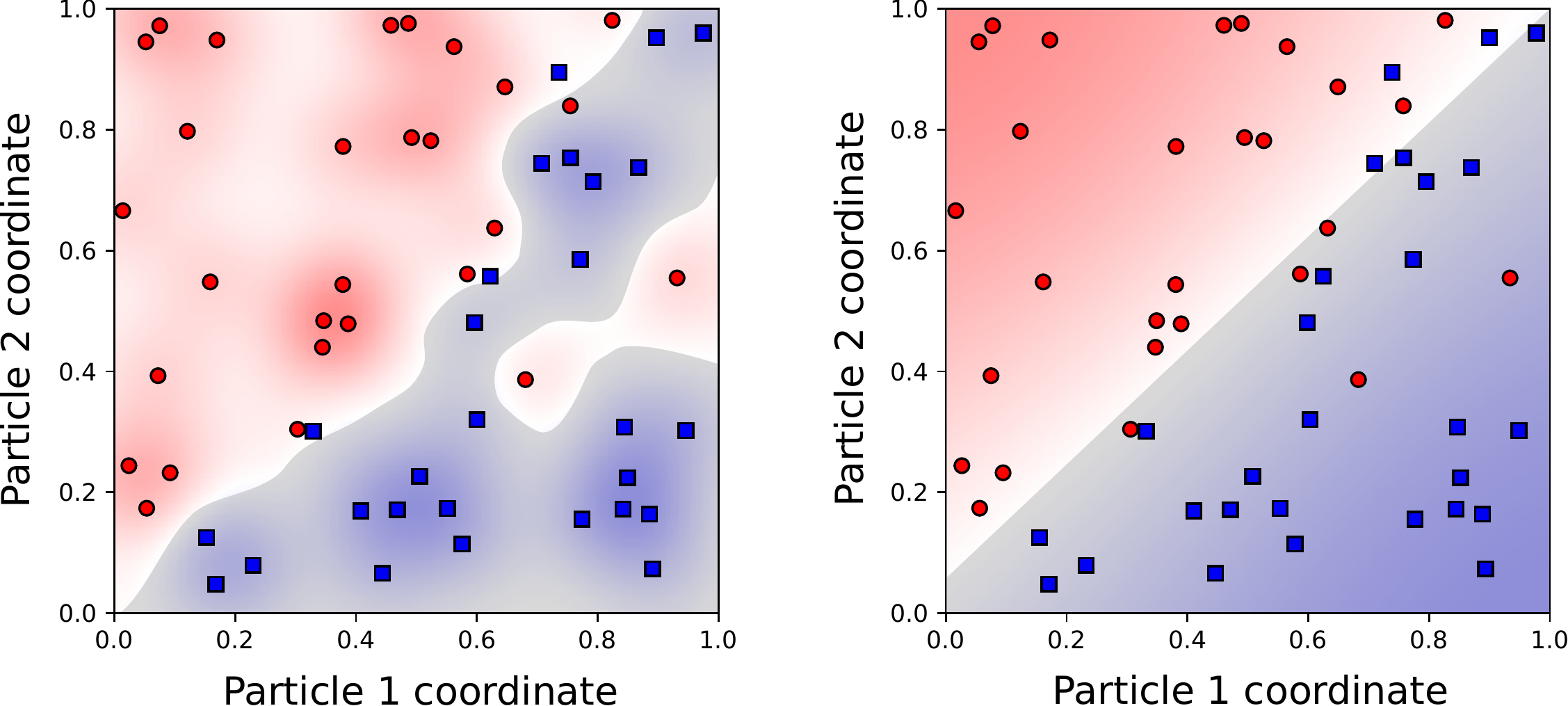}
    \caption{The effect of $\delta \tau_{\text{eff}}$ on the nodal surface of a two-fermion system in 1D. Left: $\delta\tau_{\text{eff}} = 0.1$ Right: $\delta\tau_{\text{eff}} = 1.0$. Red circles (blue squares) represent the location of positive (negative) walkers. The background is shaded according to $\psi_{D,\text{eff}}(x)$, where the nodal surface can be seen as a bright line separating the positive (red) and negative (dark blue) nodal pockets. It is clear to see that increasing $\delta \tau_{\text{eff}}$ leads to a smoother nodal surface that is closer to the analytic nodal surface at $x=y$.}
    \label{fig:tau_eff_comparison}
\end{figure}

For systems existing in one spatial dimension the nodal surface is entirely specified by the antisymmetry constraint \cite{ceperley_fermion_nodes}. As a result, fermionic methods must be tested on higher-dimensional systems, which present a significantly increased challenge. For a fixed number of walkers, the average  walker-walker separation ${\langle|x_i - x_j|\rangle_{i\neq j}}$ increases exponentially with the dimensionality of configuration space; a manifestation of the sign problem. This allows the +ve and -ve walkers more space to slip past one another and induce the bosonic collapse of the wavefunction. To mitigate this outcome we introduce an effective timestep $\delta\tau_{\text{eff}} \geq \dmct$ and enforce the nodal surface of the corresponding diffused wavefunction
\begin{equation}
\label{eq:psi_d_eff}
    \psi_{D,\text{eff}}(x) = \sum_i w_i G_D(x,x_i,\delta\tau_{\text{eff}}).
\end{equation}
By increasing $\delta\tau_{\text{eff}}$ we obtain a long-range ansatz for the nodal surface which, as before, is still free to vary in order to minimize the energy. Taking $\delta\tau_{\text{eff}} > \dmct$ can be justified on physical grounds, as the kinetic energy contribution penalises wavefunctions that fluctuate over small length scales. By increasing $\delta\tau_{\text{eff}}$, we are effectively smoothing out such fluctuations, as can be seen in Fig.\ \ref{fig:tau_eff_comparison}. However, this is still an approximation and, as such, large values of $\delta\tau_{\text{eff}}$ introduce a bias into the DMC energy which grows larger as more features of the nodal surface become unresolvable on the scale of $\delta\tau_{\text{eff}}$. In order to keep this bias as small as possible, the long range nodal surface is applied post-hoc; $\delta\tau_{\text{eff}}$ does not enter into the diffusive step or into evaluation of the functions $f_{\pm}(x)$ (see section \ref{sec:summary}).

\begin{figure}
    \centering
    \includegraphics[width=\columnwidth]{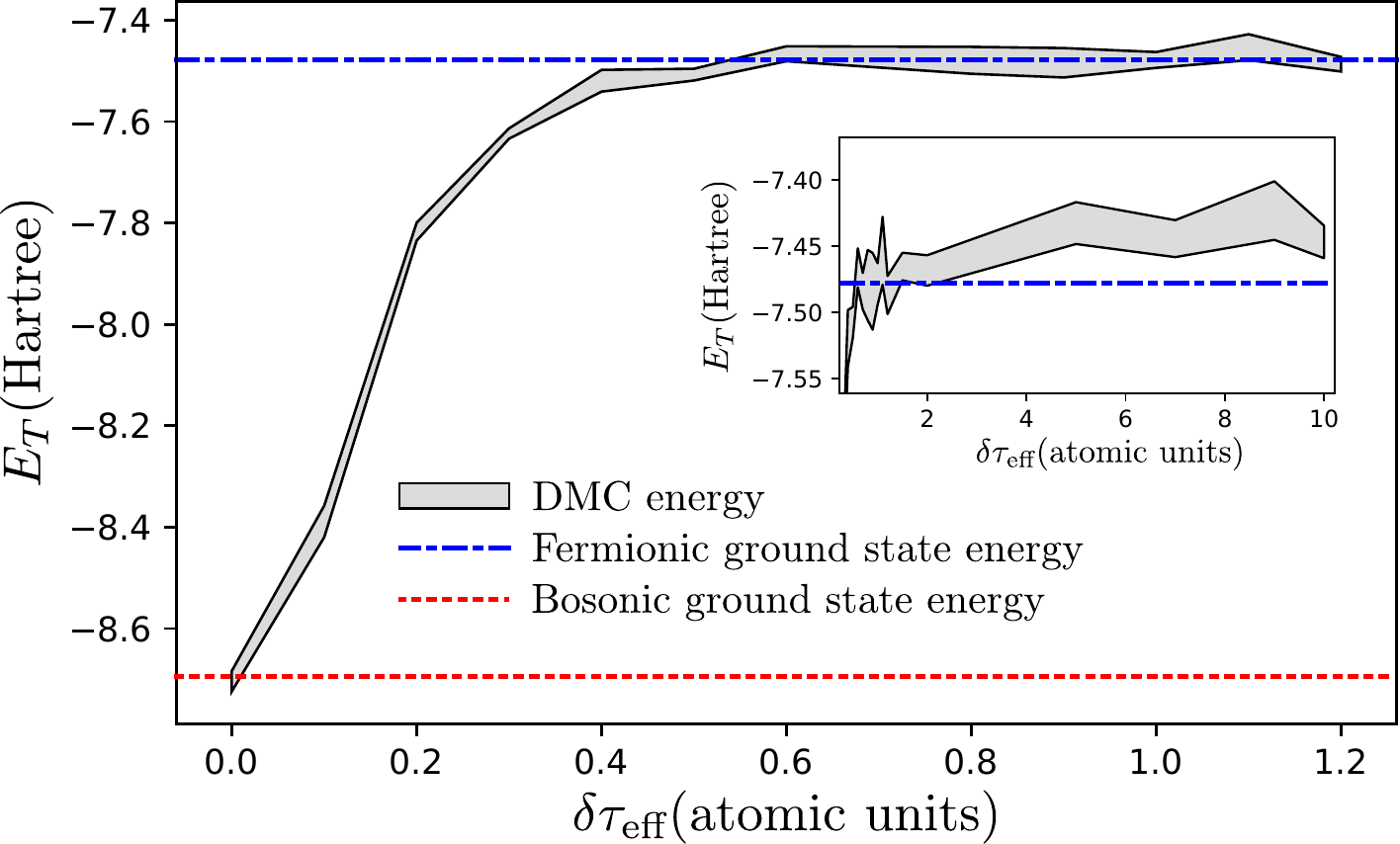}
    \caption{The DMC energy of a Lithium atom as a function of the effective timestep $\delta\tau_{\text{eff}}$ used to define the stochastic nodal surface. For each value of $\delta\tau_{\text{eff}}$, the energy was obtained from a simulation of $10^4$ walkers for $10^5$ iterations with a timestep of $10^{-3}$ atomic units. The DMC energy is shaded to $\pm$ the reblocked error. The blue dotted line is at the non-relativistic fermionic energy obtained from a Hylleraas-type expansion, accurate to within a basis set error of $< 10^{-9}$ Ha \cite{lithium_ground_state}. The inset shows the effect of increasing $\delta\tau_{\text{eff}}$ beyond sensible values.}
    \label{fig:lithium}
\end{figure}

We can see how increasing $\delta\tau_{\text{eff}}$ takes us from the bosonic ground state to the fermionic ground state of a lithium atom in Fig.\ \ref{fig:lithium}. The DMC energy plateaus at the fermionic energy as $\delta\tau_{\text{eff}}$ increases above $\sim0.6$ atomic units. We note that the resulting fermionic state is stable for long times, in contrast with transient methods such as \textit{release-node} DMC. On increasing $\delta\tau_{\text{eff}}$ beyond $\sim1.5$, we enter the regime where $\delta\tau_{\text{eff}}$ is too large to resolve the analytic nodal surface and a positive bias is introduced to the energy. This is similar to the situation in fixed-node DMC where the energy is bounded from below by the true ground state energy and variational with respect to antisymmetric trial wavefunctions.

\begin{figure}
    \centering
    \includegraphics[width=\columnwidth]{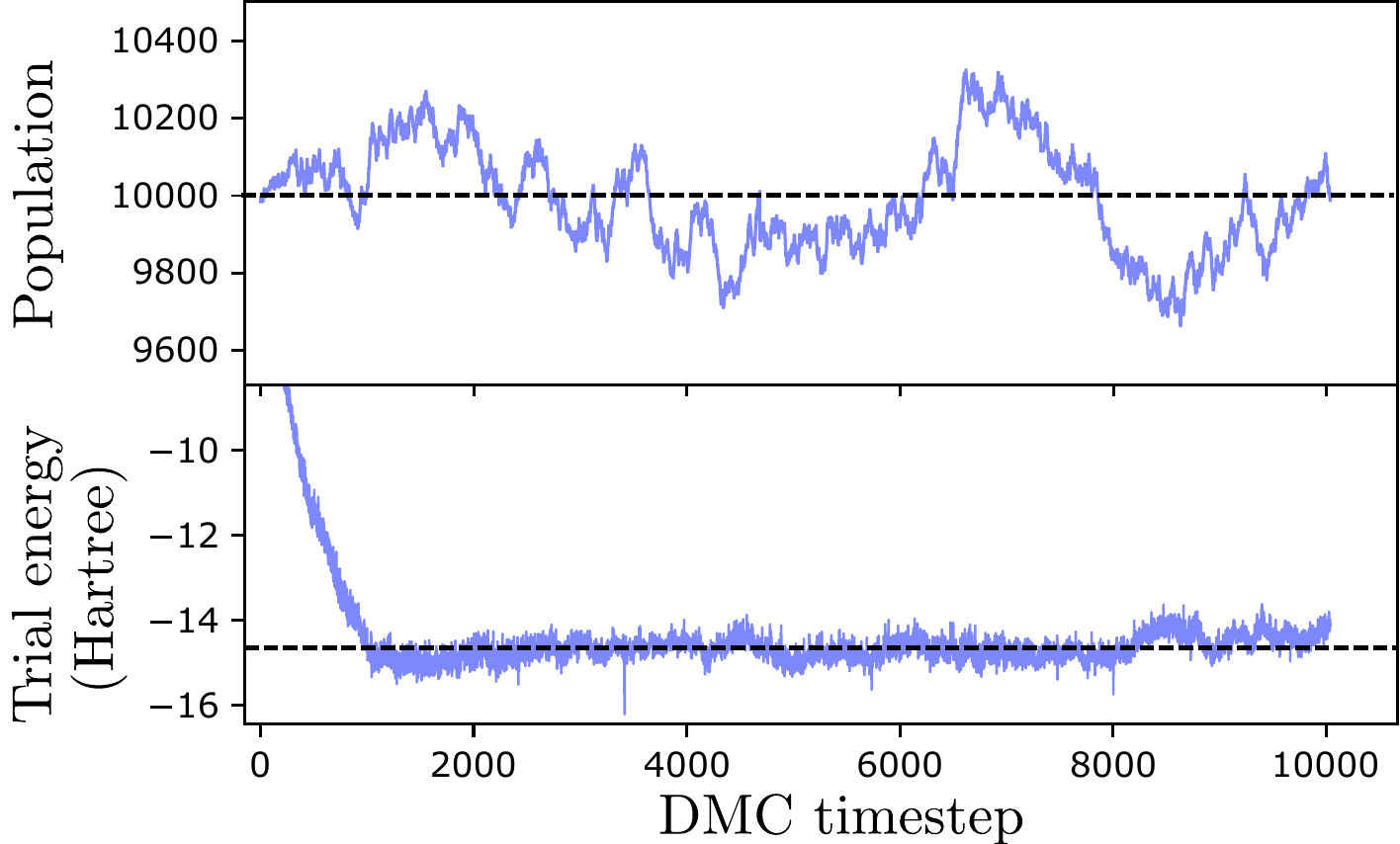}
    \caption{The evolution of a DMC calculation of a Beryllium atom. The calculation was carried out using $10^4$ walkers for $10^4$ timesteps, each of $\dmct = 10^{-3}$ atomic units. An effective timestep of $\delta\tau_{\text{eff}} = 1.35$ atomic units, derived in a preliminary calculation from Eq.\ \ref{eq:min_sep_estimator}, was used to describe the stochastic nodal surface. In the upper panel, the dashed line is at the target population. In the lower panel, the dashed line is at -14.66654 Ha, the energy obtained from a Hylleraas-type expansion, accurate to within $2\times10^{-4}$ Ha of the exact value \cite{beryllium_ground_state}. The DMC estimate of the energy is $-14.665 \pm 0.07$ Ha.}
    \label{fig:beryllium}
\end{figure}

Clearly, it would be useful to be able to identify a sensible value for $\delta\tau_{\text{eff}}$ without having to construct plots such as Fig.\ \ref{fig:lithium}. From the form of Eq.\ \ref{eq:psi_d_eff}, $\delta\tau_{\text{eff}}$ can be interpreted the range of influence of a walker on the nodal surface (see also Fig.\ \ref{fig:tau_eff_comparison}). A natural choice for it's value is then given by the expected midpoint distance between a +ve walker and it's nearest -ve neighbour:
\begin{equation}
\label{eq:min_sep_estimator}
    \delta\tau_{\text{eff}} =  \left\langle \min_{x^-} \frac{|x^+ - x^-|}{2} \right\rangle_{x^+}
\end{equation}
Where the minimization is over the positions $x^-$ of all of the negative walkers, and the average is over the positions $x^+$ of all of the positive walkers. In a preliminary calculation of the beryllium atom, the value given by Eq.\ \ref{eq:min_sep_estimator} fluctuates around $\delta\tau_{\text{eff}} = 1.35$. Carrying out an extended DMC calculation of the Beryllium ground state energy using this value for $\delta\tau_{\text{eff}}$ produces Fig.\ \ref{fig:beryllium}, from which the ground state energy is estimated as $-14.665 \pm 0.07$ Ha, well within errors of the exact value of $-14.66654\pm 2\times10^{-4}$ Ha obtained via Hylleraas-type expansions \cite{beryllium_ground_state}.

\subsection{Summary of method}
\label{sec:summary}

Combining the propagation stages explored in the preceding sections, we arrive at the following scheme
\begin{enumerate}
    \item \label{step:initialization} \textbf{Initialization} Initialise the walkers according to the procedure discussed in Sec.\ \ref{sec:initialization}.
    \item \label{step:exhange} \textbf{Exchange moves} To each walker, apply one of the exchange moves $x\rightarrow P_i x,\; w\rightarrow -w$ (or no exchange $x \rightarrow x,\; w \rightarrow w$), each with equal probability.
    \item \textbf{Diffusion} Diffuse each walker from $x \rightarrow x'$ with probability $G_D(x,x',\dmct)$.
    \item \textbf{Potential reweighting} For each walker, apply the potential reweighting $w \rightarrow wG_V(x, x', \dmct)$.
    \item \label{step:reweight} \textbf{Cancellation} Reweight positive walkers according to $w \rightarrow f_+(x')w$ and negative walkers according to $w \rightarrow f_-(x')w$ and, if $\delta\tau_{\text{eff}} > \dmct$, enforce the extended-range nodal surface of Eq.\ \ref{eq:psi_d_eff}.
    \item \label{step:branching} \textbf{Branching} Replace each walker with
    $M = \lfloor|w| + u\rfloor$ walkers each of weight $\sign(w)$ where $u$ is a uniformly-distributed random number in $[0, 1]$.
    \item \textbf{Loop} Return to step \ref{step:exhange} and repeat until expectation values have converged to the required tolerance.
\end{enumerate}
We note that steps \ref{step:exhange}-\ref{step:reweight} commute and can be applied in any order.

\section{Discussion}
\label{sec:discussion}

\subsection{Relation to previous schemes}
\subsubsection{Step \ref{step:reweight}: cancellation}

\begin{figure}[H]
\centering
    \includegraphics[width=\columnwidth]{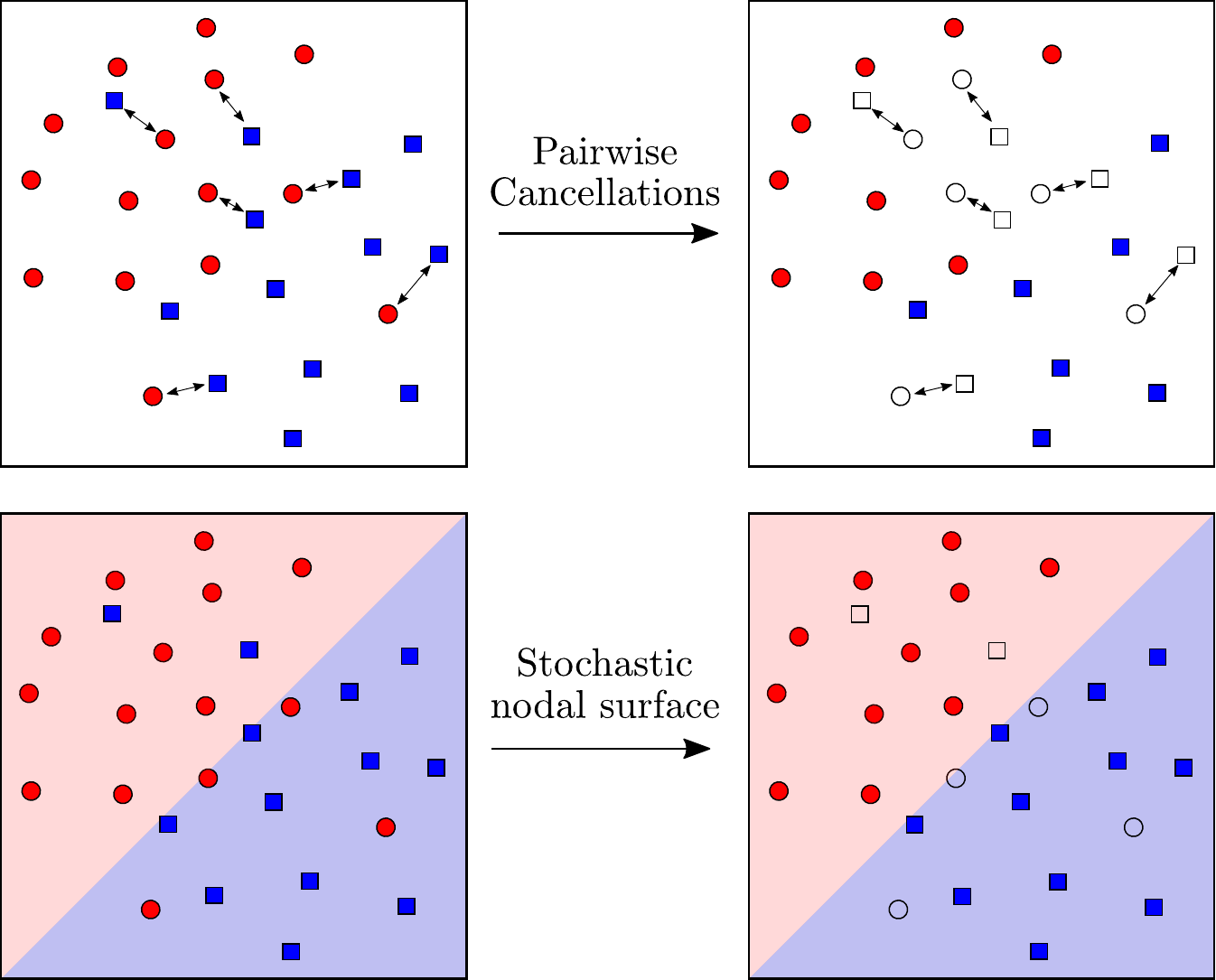}
    \caption{Schematic of cancellations via explicit pairing (upper two panels) and a stochastic nodal surface (lower two panels). Red circles represent positive walkers, blue squares represent negative walkers and empty shapes represent cancelled walkers. When using explicit pairing, walkers are first paired according to some criterion and then cancelled. This cancellation is often only partial and may take place over several iterations \cite{umrigar_observations, quantum_chemistry_by_random_walk_exact, kalos_fermion_monte_carlo}. When using a stochastic nodal surface, the diffused wavefunction is evaluated for the configuration of each walker, and any walker with the wrong sign is immediately removed from the simulation.}
    \label{fig:pairing_vs_stochastic_nodes}
    \vspace{1cm}
    \centering
    \includegraphics[width=0.75\columnwidth]{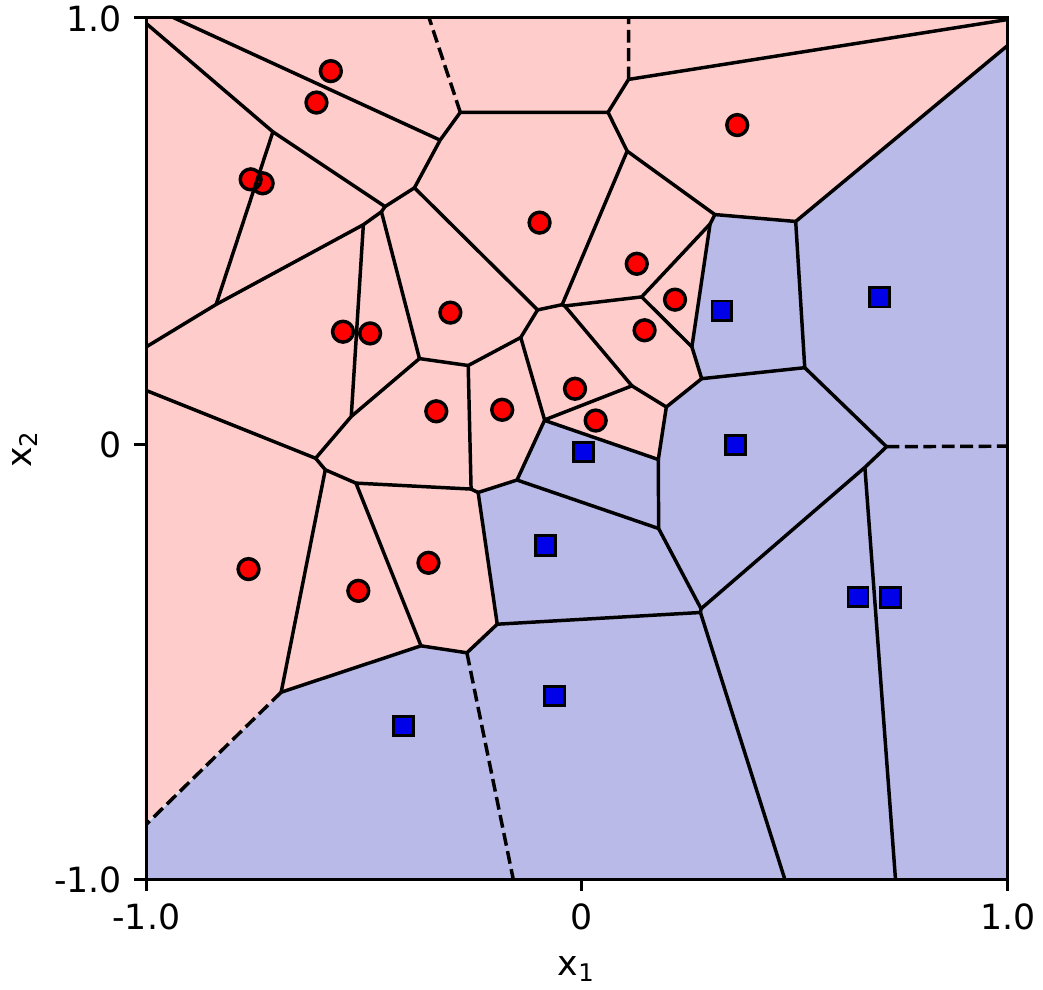}
    \caption{A Voronoi wavefunction for two non-interacting fermions in a 1D harmonic oscillator. Red circles (blue squares) represent positive (negative) walkers. The wavefunction is positive (negative) in red (dark blue) shaded regions. The emerging stochastic nodal surface at $x_1 = x_2$ can be clearly seen. Any walker crossing this surface in the next iteration will be removed from the simulation.}
    \label{fig:voronoi_wavefunction}
\end{figure}

In this work we opted not to explicitly pair walkers for cancellation and instead enforce a stochastic nodal surface defined by the entire population (see Fig.\ \ref{fig:pairing_vs_stochastic_nodes}). However, the two methods are closely related. If we consider the limiting case of cancellation between two walkers with weights $w_1 > 0$ and $w_2 < 0$ at $x_1$ and $x_2$ respectively, then Eq.\ \ref{eq:traditional_dmc_propagation} reads
\begin{equation}
\begin{aligned}
    \psi_+(x) &= w_1 G_D(x,x_1,\dmct), \\
    \psi_-(x) &= |w_2| G_D(x, x_2, \dmct).
\end{aligned}
\end{equation}
From which we can construct the cancellation function $f_\pm(x)$ according to Eq.\ \ref{eq:cancellation_function}. The reweighting given by $w\rightarrow f_\pm(x)w$ now takes the form
\begin{equation} 
\begin{aligned}
\label{eq:pairwise_cancellations}
    w_1 &\rightarrow \max(w_1 - w_2 G_D(x, x_2, \dmct)/G_D(x, x_1, \dmct), 0), \\
    w_2 &\rightarrow \max(w_2 - w_1 G_D(x, x_1, \dmct)/G_D(x, x_2, \dmct), 0).
\end{aligned}
\end{equation}
This pairwise cancellation is the same as that proposed in Refs.\ \cite{quantum_chemistry_by_random_walk_exact} and \cite{kalos_fermion_monte_carlo}. Ref. \cite{quantum_chemistry_by_random_walk_exact} goes on to show that it is possible to extend this scheme to facilitate cancellations within a collection of more than two walkers, from which Eq.\ \ref{eq:cancellation_function} can be recovered in the entire-population limit. In this work, Eq.\ \ref{eq:cancellation_function} was instead obtained directly by requiring maximal separation of the walkers into nodal pockets (see appendix \ref{app:prop_scheme}). The schemes given in Refs. \cite{quantum_chemistry_by_random_walk_exact} and \cite{kalos_fermion_monte_carlo} can therefore be thought of as limiting cases of the maximal-separation scheme when only subsets of the population are considered for cancellation. This is a sensible approximation to make if each subset consists of walkers that are near to one another, due to the limited range of the diffusive Green's function. Indeed, one could approximate $\psi_D(x)$ by only considering the $k$ nearest-neighbouring walkers to $x$, leading to
\begin{equation}
   \psi_D(x) \approx \psi_D^{(k)}(x) = \sum_{i=1}^k w_i G_D(x, x_i, \dmct)
\end{equation}
where $x_i$ are understood to be in order of increasing distance from $x$. Taking the $k=1$ case corresponds to a Voronoi tiling of configuration space, where the sign of the diffused wavefunction at $x$ is given by the sign of the nearest walker to $x$ (as shown in Fig.\ \ref{fig:voronoi_wavefunction}). The same form of nodal surface is obtained on heuristic grounds in Ref.\ \cite{antisymmetric_diffusion}, where it is shown that it produces sensible results for low-dimensional ($D<20$) configuration spaces. It is also known that the nodes of the free fermion density matrix approach that of the Voronoi wavefunction in the high temperature limit \cite{ceperley_fermion_nodes}.

\subsubsection{Step \ref{step:exhange}: Exchange moves}

It is clear that the strictly \textit{local} influence of a DMC walker is the limiting factor in describing antisymmetric wavefunctions. In this work, non-local information is introduced via the exchange moves. However, it is possible to incorporate this information implicitly in the form of the DMC walker itself. The simplest way to do this is to modify each walker to represent a set of symmetry-related points in configuration space, rather than just a single configuration:
\begin{equation}
\label{eq:antisymmetric_representation}
    \delta(x-x_i) \rightarrow \sum_{P \in \mathcal{P}} \sign(P) \delta(x-Px_i)
    \vspace{0.1cm}
\end{equation}
where $\mathcal{P}$ is the set of all fermionic permutations of the system. We can obtain this representation by imposing antisymmetry constraints for the wavefunction under any of the permutations $\mathcal{P}$, rather than the pairwise exchanges $\mathcal{E}$ that we have used thus far (see \cite{supplement}). Whilst these are equivalent problems (the exchanges $\mathcal{E}$ generate the permutations $\mathcal{P}$), they lead to different propagation schemes. The scheme arising from Eq.\ \ref{eq:antisymmetric_representation} is equivalent \cite{supplement} to the so-called \textit{second-quantized} walkers introduced in Ref.\ \cite{umrigar_observations}. As pointed out in Ref.\ \cite{umrigar_observations}, evaluating the combinatorially-many additional terms that appear in the modified form of the Green's function can be reduced from an $O(M!)$ operation to an $O(M^3)$ operation (where $M$ is the number of fermions). Whilst this is still more expensive per-iteration than stochastic sampling of the permutations via exchange-moves (where the evaluation of the greens function is an $O(M)$ operation), the additional permutations provide additional information. Employing a suitable cancellation scheme, the additional walker images introduced by the permutations can be used to increase the cancellation rate between +ve and -ve walkers \cite{supplement, umrigar_observations}. However, the efficiency of the resulting method depends strongly on how this cancellation step is implemented, and on it's effectiveness. Indeed, in the current work, evaluating Eq.\ \ref{eq:cancellation_function} for the purposes of cancellations is the rate-limiting step. For this reason, it is difficult to say in general whether permutations should be sampled directly, or via exchange-type moves.

\subsection{Scaling}
The sign problem manifests itself as an exponential increase in the computational effort required to keep the bias in the energy estimator small as the number of fermions increases. In the method described in this work, the scaling is determined by the population of walkers required to obtain a stable fermionic ground state, and how much this population can be reduced by increasing $\delta\tau_{\text{eff}}$.

\begin{figure}
    \centering
    \includegraphics[width=\columnwidth]{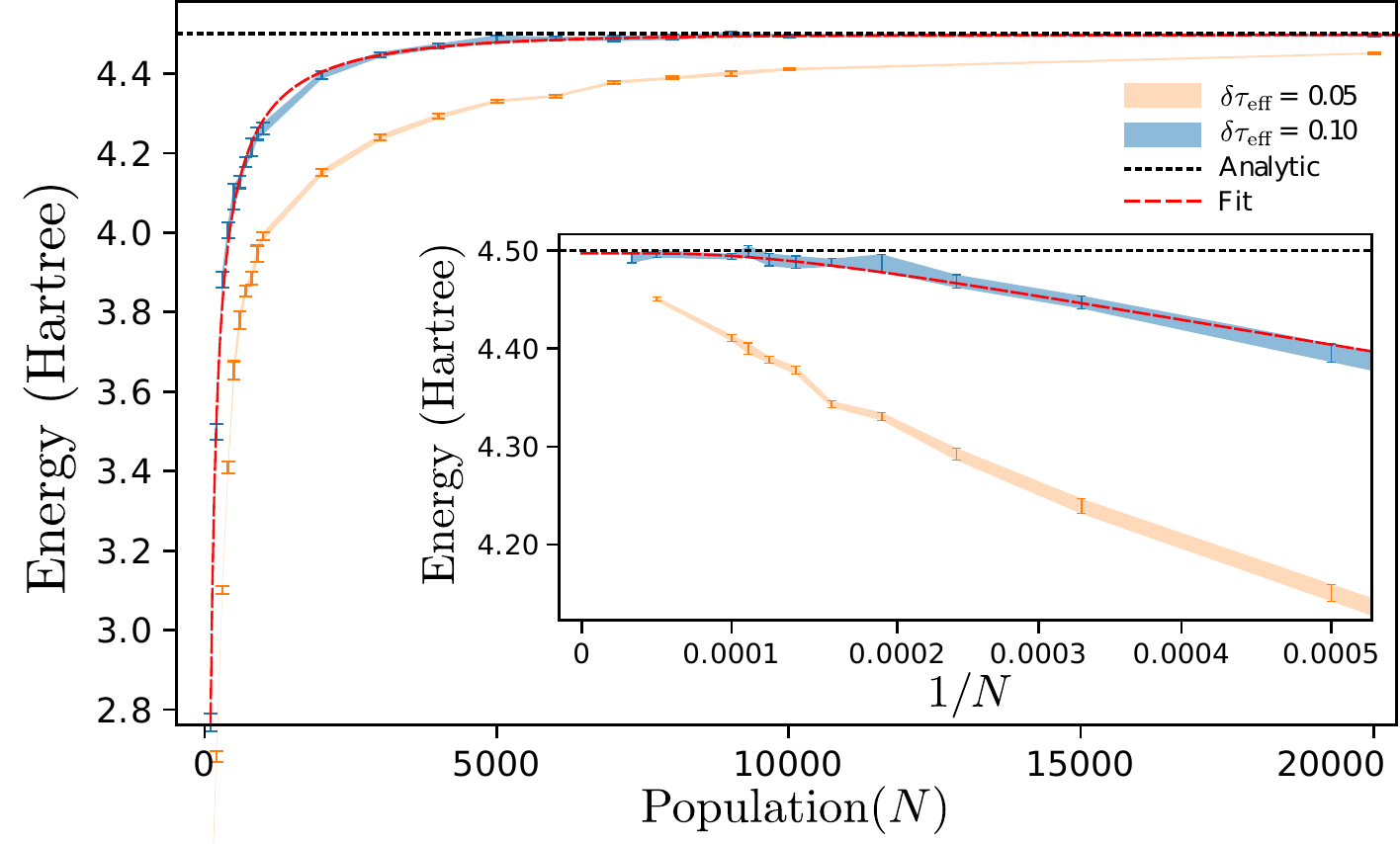}
    \caption{The DMC energy as a function of target population for three non-interacting fermions in a 1D harmonic well. The result for each population was calculated using $5\times10^4$ iterations with a timestep of $\delta\tau = 10^{-3}$ atomic units. Shown are calculations using two different values of $\delta\tau_{\text{eff}}$ (0.1 is shown in blue and converges faster than 0.05, shown in orange). The inset shows the same data plotted against the inverse population. For large populations, we found a deviation from power-law behaviour where the convergence is instead exponential. For $\delta\tau_{\text{eff}}=0.1$ the best fit converges as $N^{-0.79}\exp(-N/3542)$ and gives an energy of $4.497 \pm 0.003$ Hartree in the infinite population limit. The analytic energy is 4.5 Hartree.}
    \label{fig:e_vs_pop_3_nif}
\end{figure}

In Ref.\ \cite{fermion_monte_carlo_revisited}, it is shown that the convergence of the energy to the infinite population limit can be sped up by reducing the Bose-Fermi gap (the difference in energy between the bosonic and fermionic ground states). Typically the Bose-Fermi gap is a constant property of the Green's function being sampled. However, in the present method, the Green's function is itself constructed from the entire walker population via inter-walker cancellations in Eq.\ \ref{eq:cancellation_function} and the approximate nodal surface of Eq.\ \ref{eq:psi_d_eff}. For small populations (without a large value of $\delta\tau_{\text{eff}}$ to compensate) the cancellations due to exchanges become vanishingly probable and the we sample the bosonic dynamics of $H$ rather than fermionic dynamics of $H_X$ (this is the cause of bosonic collapse as discussed earlier). As the population increases, we approach the dynamics of $H_X$ and the Bose-Fermi gap decreases. This leads to a departure from fixed Bose-Fermi gap (power-law \cite{fermion_monte_carlo_revisited}) behaviour, as can be seen in Fig.\ \ref{fig:e_vs_pop_3_nif}. 

\begin{figure}
    \centering
    \includegraphics[width=\columnwidth]{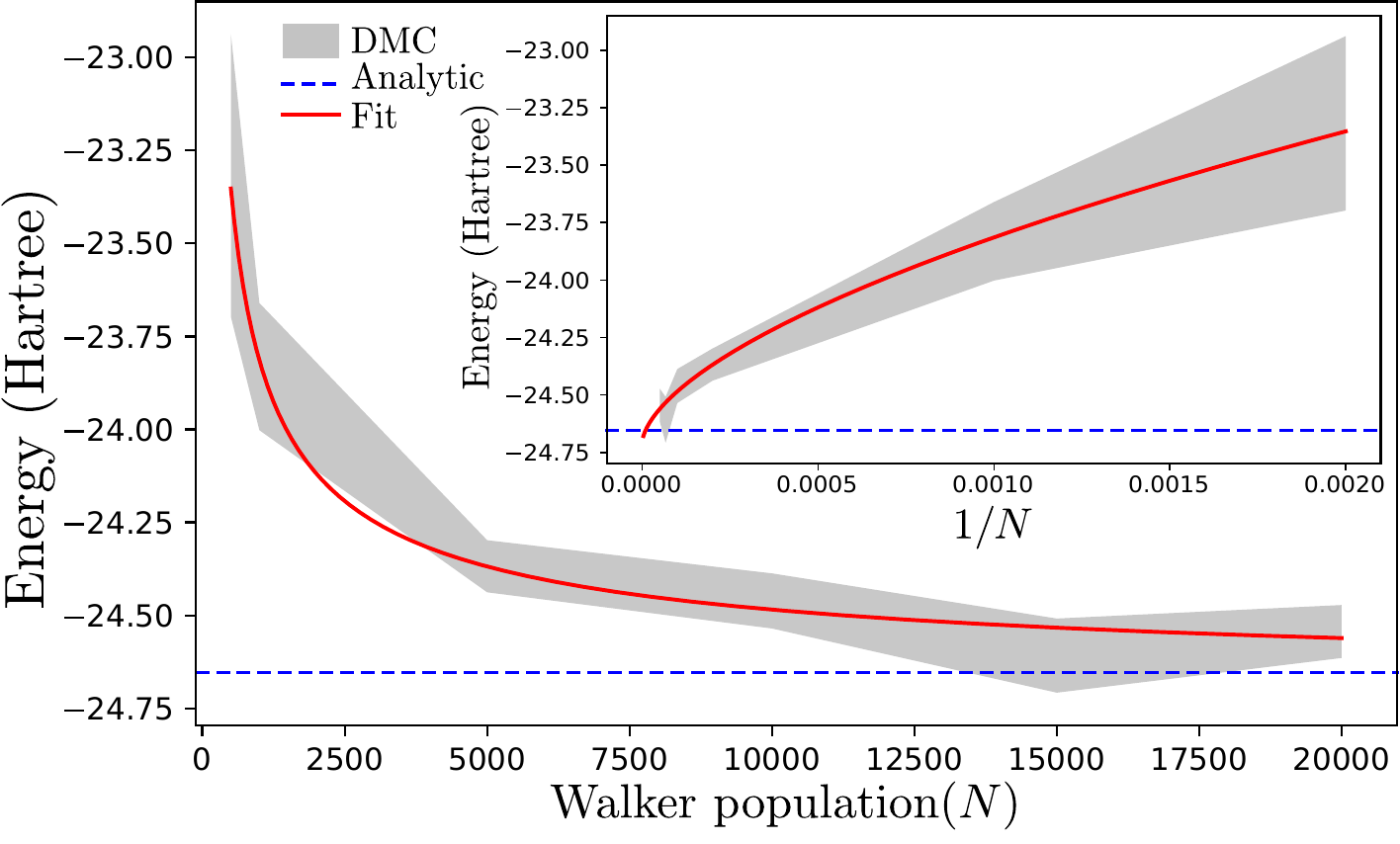}
    \caption{The DMC energy as a function of the target walker population for a Boron atom. The result for each population was calculated using $5\times10^4$ iterations with a timestep of $\delta\tau = 10^{-3}$ atomic units. $\delta\tau_{\text{eff}}$ was set to 1.35 atomic units, to facilitate comparison with our calculations of the beryllium atom. The inset shows the same data, plotted against the inverse population. The exact energy shown is at -24.65386608 $\pm 2\times10^{-9}$ Ha, which is the result obtained in the infinite-basis limit of an explicitly-correlated Gaussian basis set expansion \cite{boron_energy}. The best fit power law converges as $N^{-0.6}$ and gives an energy of $-24.67\pm0.1$ Hartree in the infinite population limit.}
    \label{fig:boron}
\end{figure}

\begin{figure}
    \centering
    \includegraphics[width=\columnwidth]{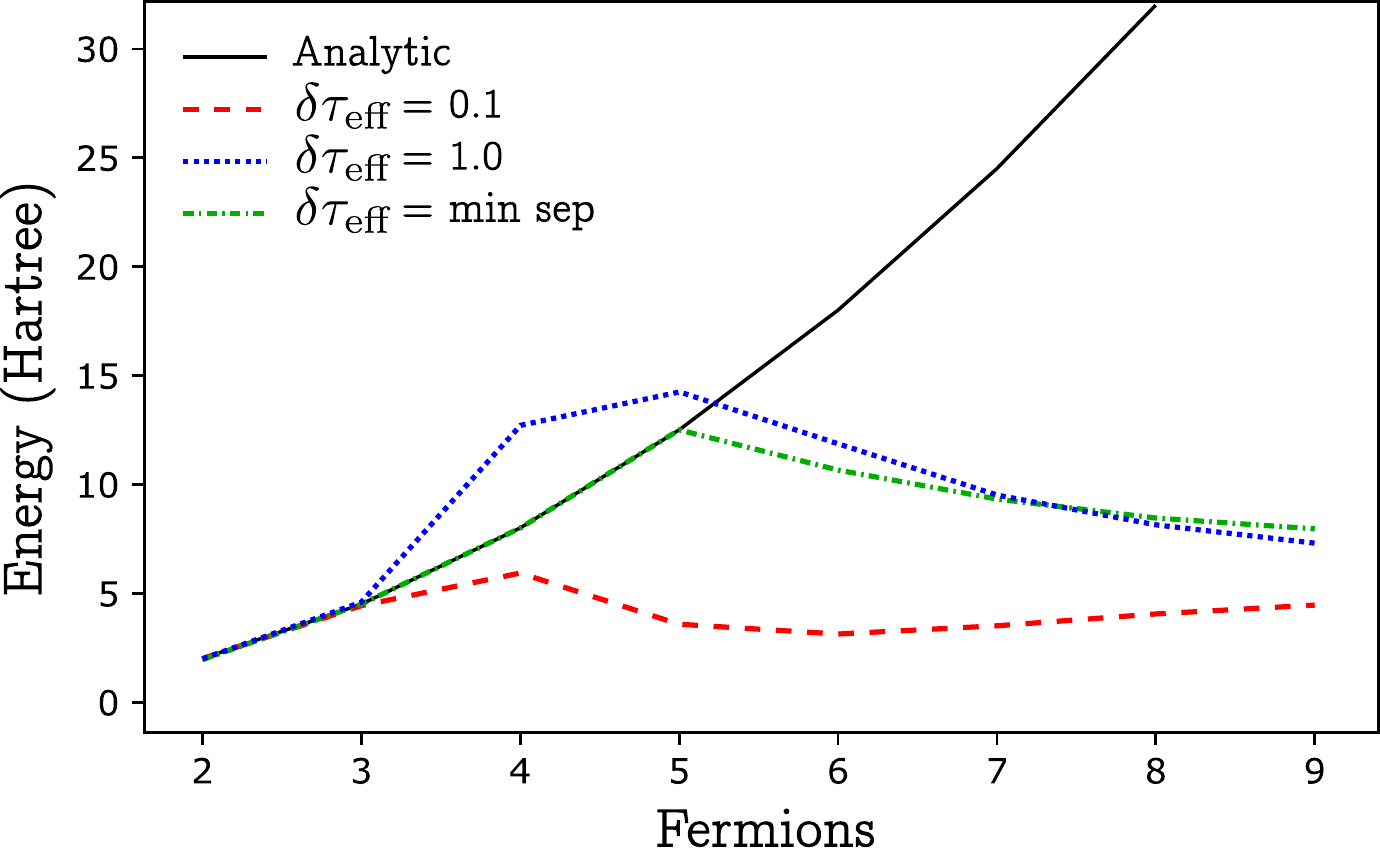}
    \caption{The energy of a system on non-interacting fermions in a 1D harmonic well, as a function of the number of fermions. The DMC calculations were carried out using 5000 walkers for $10^4$ iterations with a timestep of $\delta\tau = 10^{-3}$ atomic units. The statistical errors in the energy are smaller than the widths of the lines.}
    \label{fig:e_vs_fermions}
\end{figure}

To decrease the population required to describe a particular fermionic system we can increase the effective timestep $\delta\tau_{\text{eff}}$. The improvement in convergence as a function of population obtained by doubling $\delta\tau_{\text{eff}}$ can be seen in Fig.\ \ref{fig:e_vs_pop_3_nif}, allowing us to use around a quarter of the population for the same level of convergence. However, the amount that $\delta\tau_{\text{eff}}$ can be increased is bounded by the length scale needed to resolve the analytic nodal surface, as can be seen in Fig.\ \ref{fig:lithium} (inset), where large values of $\delta\tau_{\text{eff}}$ lead to a positive bias in the energy estimator. The optimal value of $\delta\tau_{\text{eff}}$ can be estimated using Eq.\ \ref{eq:min_sep_estimator} and, as can be seen in Fig.\ \ref{fig:e_vs_fermions}, using this value allows the description of larger systems than would otherwise be possible; we note none of the calculations in Fig.\ \ref{fig:e_vs_fermions} would lead to a fermionic result in the limit $\delta\tau_{\text{eff}} \rightarrow \delta\tau$. However, as the number of fermions increases, the number of walkers required to describe the nodal surface eventually also increases, regardless of the choice of $\delta\tau_{\text{eff}}$ (forcing one to perform infinite population extrapolations as in Figs.\ \ref{fig:e_vs_pop_3_nif} and \ref{fig:boron}). This can also be seen in Fig.\ \ref{fig:e_vs_fermions} where, on increasing the number of fermions, (partial \footnote{Cancellations still occur, and increase the energy estimator, just not at a sufficient rate to stabilise the fermionic ground state.}) bosonic collapse eventually occurs leading to a large underestimation of the energy. As a result, even though increasing $\delta\tau_{\text{eff}}$ enables a finite set of walkers to describe larger systems than would otherwise be possible, it cannot be increased fast enough with system size to completely overcome the sign problem.

\section{Conclusion}

We have constructed a scheme for fermionic diffusion Monte Carlo that makes no reference to a trial wavefunction. We have shown how the resulting propagation scheme can be interpreted as the formation of a \textit{stochastic} nodal surface, which is free to vary and minimize the energy. We go on to derive a diffusion scheme that maximally stabilizes the nodal surface and show that stable fermionic ground states for simple harmonic systems and light atoms can be obtained. We have demonstrated that the number of walkers required to resolve the nodal surface can be reduced, without introducing significant bias, by introducing an approximate long-range influence on the nodal surface and have provided a method for estimating a sensible choice for the associated parameter ($\delta\tau_{\text{eff}}$). Extending the method to employ a guiding wavefunction for the purposes of importance sampling and to allow the use of projection estimators should allow the study of larger systems, but the sign problem still persists for the method in its current form. We hope that methods based on the constrained-optimization formalism of DMC introduced in this work will enable studies to improve the understanding of nodal surfaces in electronic wavefunctions. We also plan to apply this method to the study of exchange and correlation in periodic systems, with the ultimate goal of generating exchange-correlation functionals for DFT calculations that do not depend on a choice of trial wavefunction at the DMC level. An open-source \textsc{C++} implementation of the methods in this work is available \cite{github}.

\appendix
\section{The Green's function of \texorpdfstring{$H_X$}{Hx}}
\label{app:greens_function}

For small timesteps, we derive the form of the Green's function
\begin{equation}
   G(x,x',\dmct) = \bra{x} \exp{(-\dmct H_X)} \ket{x'}.
\end{equation}
Writing $H = T + V$ where $T$ is the kinetic energy operator and $V$ is a local potential we can apply the Suzuki-Trotter \cite{trotter_suzuki, qmc_needs} expansion of the Green's function to obtain
\begin{equation}
\begin{aligned}
\label{eq:greens_function_intermediate}
    &G(x, x', \dmct) \approx  G_V(x, x', \dmct) \times  \\
    &\bra{x} \exp(-\dmct[T + \sum_i \mu_i(x) (P_i+1)]) \ket{x'}.
\end{aligned}
\end{equation}
For small timesteps, the exponential can be factorized, allowing us to write
\begin{equation}
\begin{aligned}
    \bra{x} \exp(-\dmct T)[1 - &\sum_i \dmct \mu_i(x)(P_i+1)] \ket{x'} \\
    \approx \big[1 - &\sum_i \dmct \mu_i(x')\big] G_D(x,x',\dmct) \\
    -&\sum_i \dmct \mu_i(P_ix') G_D(x,P_ix',\dmct).
\end{aligned}
\end{equation}
Noting that
\begin{equation}
\begin{aligned}
   |x-P_ix'| &= |P_i(x - P_ix')| = |P_ix - x'| \\ 
   &\implies G_D(x, P_ix', \dmct) = G_D(P_ix, x', \dmct)
\end{aligned}
\end{equation}
and that, because $P_i$ corresponds to exchanging identical particles,
\begin{equation}
\begin{aligned}
    V(P_ix) &= V(x) \implies G_V(P_ix, x', \dmct) = G_V(x, x', \dmct),
\end{aligned}
\end{equation}
we can finally write the Green's function as
\begin{equation}
\label{eq:greens_function_final}
\begin{aligned}
    G(x,x',\dmct) &= \bigg[\left(1 - \sum_i \dmct \mu_i(x')\right)-\sum_i\dmct\mu_i(P_ix')P_i \bigg]
    \\ & \times G_V(x,x',\dmct)G_D(x,x',\dmct)
\end{aligned}
\end{equation}
where $P_i$ now acts on the unprimed (pre-propagation) coordinates.

\section{Maximum-separation propagation}

\label{app:prop_scheme}
In order to encourage the formation of nodal pockets, we seek the form of $P_\pm(x)$ that maximizes the expected separation of +ve and -ve walkers, given by
\begin{equation}
    \langle|x_+ - x_-|\rangle = \int P_+(x_+)P_-(x_-)|x_+ - x_-| dx_+ dx_-.
\end{equation}
This is equivalent to extremizing
\begin{equation}
\begin{aligned}
    \mathcal{S} = &\int S_+^2(x_+)S_-^2(x_-)|x_+ - x_-|  dx_+ dx_- \\
    + &\int \lambda(x)[S_+^2(x)-S_-^2(x)-\psi_D(x)]dx
\end{aligned}
\end{equation}
with respect to $S^2_\pm(x) = P_\pm(x)$ (introduced to ensure $P_\pm(x) \geq 0$) and the Lagrange multiplier $\lambda(x)$ which enforces the constraint $\psi_D(x) = P_+(x) - P_-(x)$. Extremization of $\mathcal{S}$ leads to
\begin{align}
    \frac{\delta \mathcal{S}}{\delta S_+(y)} = \int 2S_+(y)S_-^2(z)|z-y|dz + 2S_+(y)\lambda(y) \overset{!}{=} 0, \label{eq:s_plus_extremize} \\
    \frac{\delta \mathcal{S}}{\delta S_-(y)} = \int 2S_-(y)S_+^2(z)|z-y|dz - 2S_-(y)\lambda(y) \overset{!}{=} 0. \label{eq:s_minus_extremize}
\end{align}
Now, if we assume that both $S_+(y) \neq 0$ and $S_-(y) \neq 0$, Eqs.\ \ref{eq:s_plus_extremize} and \ref{eq:s_minus_extremize} read
\begin{align}
    \frac{1}{2S_+(y)}\frac{\delta \mathcal{S}}{\delta S_+(y)} = \int S_-^2(z)|z-y|dz + \lambda(y) = 0, \\
    \frac{1}{2S_-(y)}\frac{\delta \mathcal{S}}{\delta S_-(y)} = \int S_+^2(z)|z-y|dz - \lambda(y) = 0.
\end{align}
Adding these equations gives
\begin{equation}
    \int [S_+^2(z)+S_-^2(z)]|z-y|dz = 0 \implies S_+^2(z)+S_-^2(z) = 0,
\end{equation}
a contradiction. This means that at most one of $S^2_+(y) = P_+(x)$ and $S^2_-(y) = P_-(x)$ is non-zero (i.e the distributions of +ve walkers and -ve walkers are mutually exclusive). Combined with the condition $\psi_D(x) = P_+(x) - P_-(x)$, we must have
\begin{equation}
    P_\pm(x) = 
    \begin{cases} 
          |\psi_D(x)| & \text{if}\; \sign(\psi_D(x)) = \pm 1, \\
          0           & \text{otherwise}.
    \end{cases}
    \vspace{0.25cm}
\end{equation}
Note that this derivation does not depend on the form of $\psi_D(x)$. It also results in the same distributions $P_\pm(x)$ for any measure of separation that is symmetric in $x_+$ and $x_-$, not just $|x_+ - x_-|$.

\section*{Acknowledgements}
M.H.\ would like to thank his supervisor Richard Needs for the academic freedom to persue side projects such as this as well as Nick Woods and Cyrus Umrigar for helpful discussions. He also acknowledges the EPSRC Centre for Doctoral Training in Computational Methods for Materials Science for funding under grant number EP/L015552/1.

\bibliography{references.bib}

\section{Supplementary information for \textit{stochastic nodal surfaces in quantum Monte Carlo calculations}}

\subsection{Variation of $\mathcal{L}$}
We look for extrema of
\begin{equation}
    \mathcal{L}[\psi] = E_T + \bra{\psi} H_X \ket{\psi}
\end{equation}
with respect to variation of $\psi$ and $\psi^*$. Variations in $\psi^*$ are straightforward
\begin{equation}
\begin{aligned}
    \mathcal{L}[\psi^* + \delta\psi^*] &= E_T + \int (\psi^* + \delta\psi^*) H_X \psi dx \\
    &= \mathcal{L}[\psi] + \int \delta\psi^* H_X \psi dx \\
    &\underbrace{\overset{!}{=} \mathcal{L}[\psi] \;\forall\; \delta\psi^*}_{\text{Extremization}} \implies H_X \psi = 0.
\end{aligned}
\end{equation}
Variations in $\psi$ are more involved
\begin{equation}
\begin{aligned}
\label{eq:variation_in_psi}
    \mathcal{L}[\psi + \delta\psi] &= E_T + \int \psi^* H_X (\psi + \delta\psi) dx \\
    &= \mathcal{L\mathcal{L}[\psi] +}[\psi] + \int \psi^* H_X \delta\psi dx\\
    &= \mathcal{L}[\psi] + \int \psi^* \left[T + V + \sum_i \mu_i(1+P_i)\right] \delta\psi.
\end{aligned}
\end{equation}
We can shift the kinetic term to instead operate on $\psi^*$ by using integration by parts twice:
\begin{equation}
\begin{aligned}
    &\int \psi^* \frac{\partial^2 \delta\psi}{\partial x_i^2}dx \\
    &= \cancelto{0}{\left[\psi^*\frac{\partial\delta\psi}{\partial x_i}\right]_\infty} - \int \frac{\partial \psi^*}{\partial x_i} \frac{\partial \delta\psi}{\partial x_i} dx \\
    &= -\cancelto{0}{\left[\frac{\partial \psi^*}{\partial x_i} \delta\psi \right]_\infty} + \int \frac{\partial^2 \psi^*}{\partial x_i^2} \delta\psi dx
\end{aligned}
\end{equation}
where we have assumed that $\psi \rightarrow 0$ as $|x| \rightarrow \infty$ to cancel the boundary terms. We can also act with permutation operators to the left within the integral because
\begin{equation}
\begin{aligned}
    &  \int f(x) P_i g(x) dx \\
    &= \int f(x) g(P_i x) dx \\
    & \hspace{0.5cm} \text{let} \; z = P_i x \rightarrow \\
    &= \int f(P_i z) g(z) dz \\
    & \hspace{0.5cm} \text{relabel} \; z \rightarrow x \\
    &= \int f(P_i x) g(x) dx.
\end{aligned}
\end{equation}
Putting this together we can write
\begin{equation}
     \mathcal{L}[\psi + \delta\psi] = \mathcal{L}[\psi] + \int \delta\psi \left[T + V + \sum_i (1+P_i) \mu_i\right] \psi^* dx
\end{equation}
where the permutation operators act to the right. Note that $\mu_i$ now appears after the permutation operators. If we assume $\mu_i$ is symmetric with respect to permutations (in the main text it is a constant because we take the exchanges to be equiprobable) then we can pull it back through the permeation operators and write
\begin{equation}
\begin{aligned}
    \mathcal{L}[\psi + \delta\psi] &= \mathcal{L}[\psi] + \int \delta \psi H_X \psi^* dx \\
    &\underbrace{\overset{!}{=} \mathcal{L}[\psi] \;\forall\;  \delta\psi}_{\text{Extremization}} \implies H_X \psi^* = 0.
\end{aligned}
\end{equation}

\subsection{Second-quantized walkers}
Consider if, instead of solving the optimization problem with the wavefunction constrained to be antisymmetric w.r.t pairwise fermionic exchanges $\mathcal{E}$, we were to constrain the wavefunction to pick up the sign of any of the fermionic \textit{permutations} $\mathcal{P}$. The constraints that must be satisfied are then
\begin{equation}
    \psi(x) = \sign(P)\psi(Px) \; \forall \; P \in \mathcal{P}.
\end{equation}
These constraints are equivalent to imposing the exchange constraints
\begin{equation}
    \psi(x) = -\psi(Ex) \; \forall \; E \in \mathcal{E},
\end{equation}
as any element of $\mathcal{P}$ can be obtained as a combination of exchanges from $\mathcal{E}$. Similarly to the exchange case, permutations also result in an effective Hamiltonian:
\begin{equation}
    H_P = H - E_T + \sum_{P\in\mathcal{P}} \mu_P(x) (1 - \sign(P)P).
\end{equation}
The Green's function for this Hamiltonian is given by
\begin{equation}
   G_P(x,x',\dmct) = \bra{x} \exp{(-\dmct H_P)} \ket{x'}.
\end{equation}
Following the derivation for $H_X$, for small timesteps $\dmct$ we have
\begin{equation}
\begin{aligned}
    G_P(x, x', \dmct) &\approx G_V(x, x', \dmct) \times \\
    \bra{x} \exp(-\dmct T)&\left[1 - \sum_{P\in\mathcal{P}} \dmct \mu_P(x)(1-\sign(P)P)\right] \ket{x'} \\
    &= G_V(x, x', \dmct) \times
    \\ &\bigg[ \underbrace{\left(1 - \sum_{P\in\mathcal{P}} \dmct \mu_P(x')\right)}_{\mathcal{N}_P(x')} \underbrace{G_D(x, x', \dmct)}_{\text{Diffusion from }x'\rightarrow x} 
    \\+ &\sum_{P\in\mathcal{P}} \underbrace{\dmct \mu_P(Px')}_{\mathcal{X}_p(x')} \sign(P) \underbrace{G_D(x, Px', \dmct)}_{\text{Diffusion from } Px' \rightarrow x}\bigg].
\end{aligned}
\end{equation}
Choosing the $\mu_P(x)$'s such that $\mathcal{N}_P = 0$ and $\mathcal{X}_P(x')$ is constant (similarly to what we do for simplicity in the $H_X$ case), and noting that for a matrix $A$ with entries $A_{i,j}$
\begin{equation}
    \det(A) = |A| = \sum_{P\in\mathcal{P}} \sign(P) \prod_i A_{i,P_i},
\end{equation}
we can obtain the form of the Green's function proposed in Eq. 13 of Ref. \cite{umrigar_observations}:
\begin{equation}
\label{eq:second_quantized_greens_function}
\begin{aligned}
    &G_P(x,x',\dmct) = \\
    &G_V(x,x',\dmct) \begin{vmatrix}
        g(x_1, x'_1) & g(x_1, x'_2) & \dots & g(x_1, x'_n) \\
        g(x_2, x'_1) & g(x_2, x'_2) & \dots & g(x_2, x'_n) \\
        \vdots       &              &       & \vdots       \\
        g(x_N, x'_1) & g(x_N, x'_2) & \dots & g(x_N, x'_N) 
    \end{vmatrix}
\end{aligned}
\end{equation}
where
\begin{equation}
    g(x_i,x'_j) = \frac{1}{\sqrt{2\pi\dmct}} \exp\left(-\frac{(x_i-x'_j)^2}{2\dmct}\right)
\end{equation}
takes single-particle coordinates from the primed and unprimed configurations as arguments.

\begin{figure}
    \centering
    \includegraphics[width=\columnwidth]{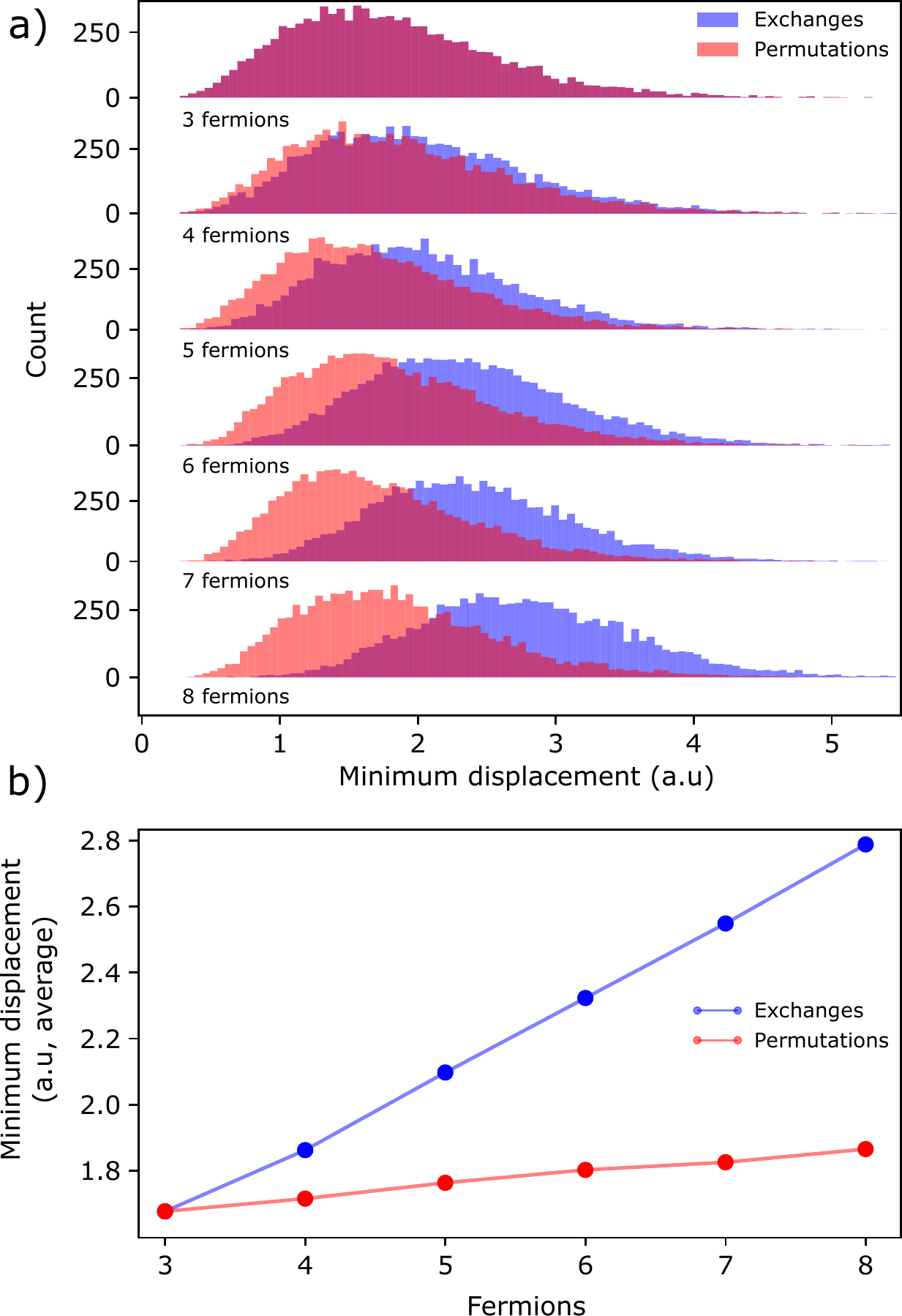}
    \caption{a) The distribution of $D$, given in Eq.\ \ref{eq:d_distribution}. b) The average of the distributions in a), plotted against fermion number, showing the linear trend.}
    \label{fig:ex_vs_perm}
\end{figure}

The propagation of walkers according to Eq.\ \ref{eq:second_quantized_greens_function} is an alternative method to the propagation using exchange moves. It can be interpreted as the propagation of a collection of \textit{second-quantized} walkers, each consisting of $N!$ symmetry-related delta-function walkers. Due to it's detrimental form Eq.\ \ref{eq:second_quantized_greens_function} can be evaluated in $O(N^3)$ time, rather than $O(N!)$ time \cite{umrigar_observations}. This is, however, still more expensive than the exchange-moves scheme where the Green's function can be evaluated in $O(N)$ time. However, the additional information contained within the second-quantized walkers could overcome this shortcoming. The utility of the additional permutations can be probed by considering the scope for additional cancellations. In Fig.\ \ref{fig:ex_vs_perm}(a), we plot the distribution of the minimum separation obtainable by applying permutations to two walkers, each distributed according to a Gaussian with $\sigma = 1$ atomic unit. Mathematically speaking, we are plotting the distribution of the distance $D$ given by
\begin{equation}
\label{eq:d_distribution}
    D = \min_{P \in S} |x - Py| \;\text{where}\; x, y \sim \mathcal{N}(\mu=0, \sigma=1)
\end{equation}
where $S$ is either the set of exchanges $\mathcal{E}$ or the set of permutations $\mathcal{P}$. Ideally, this distance would be as small as possible to facilitate cancellations between oppositely-signed walkers. In Fig.\ \ref{fig:ex_vs_perm}(b), we plot the average value of $D$ against the number of fermions in the system. This distance increases linearly as the fermion count increases, resulting in an exponentially decreasing possibility for cancellation; a manifestation of the sign problem. This increase is slower when using the full set of permutations $\mathcal{P}$, because it affords us more freedom in the permuted configurations ($|\mathcal{P}| > |\mathcal{E}|$). The choice of whether to use exchanges or permutations depends on how well this additional freedom can be exploited in an algorithmic setting.

\subsection{Form of cancellation functions}
Here, we show that
\begin{equation}
    P_\pm(x) = 
    \begin{cases} 
          |\psi_D(x)| & \text{if}\; \sign(\psi_D(x)) = \pm 1, \\
          0          & \text{otherwise}.
    \end{cases} \label{eq:prop_scheme}
\end{equation}
can be written as
\begin{equation}
\label{eq:propagation_factorisation}
    P_\pm(x) = \psi_\pm(x)f_\pm(x)
\end{equation}
where 
\begin{equation}
\begin{aligned}
\label{eq:psi_plus_minus}
    \psi_+(x) &= \sum_{w_i > 0} w_i G_D(x, x_i(\tau),\dmct) \geq 0, \\
    \psi_-(x) &= \sum_{w_i < 0} |w_i| G_D(x, x_i(\tau),\dmct) \geq 0
\end{aligned}
\end{equation}
and
\begin{equation}
    f_\pm(x) = \max\left(1 - \psi_\mp(x)/\psi_\pm(x),\; 0\right) \in [0,1].
\end{equation}
Examining the form of $\psi_D(x)$, we have
\begin{align}
    \psi_D(x) = &\psi_+(x) - \psi_-(x)\\
    \label{eq:f_plus_equality}
    = &\psi_+(x)\underbrace{\left[1-\frac{\psi_-(x)}{\psi_+(x)}\right]}_{=\;f_+(x)\;\text{if}\;\psi_D(x) \;>\; 0} \\
    \label{eq:f_minus_equality}
    = -&\psi_-(x)\underbrace{\left[1-\frac{\psi_+(x)}{\psi_-(x)}\right]}_{=\;f_-(x)\;\text{if}\;\psi_D(x) \;<\; 0}
\end{align}
and
\begin{equation}
\begin{aligned}
    0 \leq f_+(x) \leq 1 &\;\;\text{if}\;\;\psi_D(x) > 0, \\
    0 \leq f_-(x) \leq 1 &\;\;\text{if}\;\;\psi_D(x) < 0. \\
\end{aligned}
\end{equation}
Using Eq.\ \ref{eq:f_plus_equality} when $\psi_D(x) > 0$ and Eq.\ \ref{eq:f_minus_equality} when $\psi_D(x) < 0$ allows us to combine both into the compact form of Eq.\ \ref{eq:propagation_factorisation}. The form of Eq.\ \ref{eq:propagation_factorisation} then allows us to interpret $f_\pm(x)$ as a weight cancellation function. In certain limits, this function leads to cancellation-based schemes proposed in the past \cite{antisymmetric_diffusion, quantum_chemistry_by_random_walk_exact, supplement}. The prefactor of $f_\pm(x)$ in Eq. \ref{eq:propagation_factorisation} is simply the diffused wavefunction for the corresponding sign, $\psi_\pm(x)$. This means we can diffuse a walker with weight $w$ from $x \rightarrow y$ normally according to $G_D(y,x,\dmct)$ so long as we then apply the weight update
\begin{equation}
\label{eq:nodal_surface_enforce}
    w \rightarrow
    \begin{cases} 
      f_+(y)w & \text{if}\; \psi_+(y) > \psi_-(y) \And w > 0, \\
      f_-(y)w & \text{if}\; \psi_+(y) < \psi_-(y) \And w < 0, \\
      0 & \text{otherwise}.
   \end{cases}
\end{equation}
Where we evaluate $\psi_\pm(x, \tau+\dmct)$ directly via Eq.\ \ref{eq:psi_plus_minus}.

\bibliography{references.bib}

\end{document}